\documentclass[journal,10pt]{IEEEtran}

\usepackage{cite}
\usepackage{graphicx}
\usepackage{amsmath}
\usepackage{array}
\usepackage{mdwmath}
\usepackage{amssymb}
\usepackage{mdwtab}
\usepackage{stfloats}
\usepackage[tight,footnotesize]{subfigure}
\usepackage{amsmath,amsthm}
\usepackage{threeparttable}
\usepackage{color}
\usepackage{url}
\usepackage[ruled, linesnumbered]{algorithm2e}
\usepackage{multicol}
\usepackage{amssymb}
\usepackage{epstopdf}
\usepackage{bm}

\usepackage{hyperref}

\begin{document}
	
\title{Exploiting Movable Antennas in NOMA Networks: Joint Beamforming, Power Allocation and Antenna Position Optimization}

\author{Yufeng Zhou, Wen Chen, \IEEEmembership{Senior Member,~IEEE}, Qingqing Wu, \IEEEmembership{Senior Member,~IEEE}, Xusheng Zhu,\\ Zhendong Li, Kunlun Wang, Qiong Wu
	
\thanks{Y. Zhou, W. Chen, and Q. Wu are with the Department of Electronic Engineering, Shanghai Jiao Tong University, Shanghai 200240, China (e-mail: ereaked@sjtu.edu.cn; wenchen@sjtu.edu.cn; qingqingwu@sjtu.edu.cn).
	
	X. Zhu is affiliated with the Department of Electronic and Electrical
	Engineering, University College London, WC1E 7JE, United
	Kingdom (email: xusheng.zhu@ucl.ac.uk).
	
	Z. Li is with the School of Information and Communication Engineering, Xi’an Jiaotong University, Xi’an 710049, China (email:
	lizhendong@xjtu.edu.cn)
	
	K.Wang is with the School of Communication and Electronic
	Engineering, East China Normal University, Shanghai 200241, China (e-mail:
	klwang@cee.ecnu.edu.cn).
	
	Qiong Wu are with the School of Internet of Things
	Engineering, Jiangnan University, Wuxi 214122, China (qiongwu@jiangnan.edu.cn).
	}
}
	
\markboth{}
{}
\maketitle

\begin{abstract}
	This paper investigates the movable antenna (MA)-assisted downlink non-orthogonal multiple access (NOMA) network to maximize system throughput. In the considered scenario, both the base station (BS) and users are equipped with MA, and a predetermined successive interference cancellation (SIC) decoding order is adopted. Based on the field-response channel model, we formulate a complex, non-convex problem to jointly optimize the BS beamforming, power allocation, and MA positions at both the transmitter and receivers. To address this, we propose an efficient algorithm based on an alternating optimization (AO) framework, which decomposes the original problem into three distinct subproblems. By employing sequential parametric convex approximation (SPCA) and successive convex approximation (SCA) techniques, the non-convex constraints within each subproblem are transformed into tractable. This methodology ensures the algorithm converges to a stable, locally optimal solution. Numerical results validate that the proposed system, which fully exploits the degrees of freedom from antenna mobility at both ends, significantly outperforms benchmarks in terms of throughput.
\end{abstract}


\section{Introduction}

As research on fifth-generation (5G) wireless communications reaches maturity, the focus is gradually shifting toward the development of sixth-generation (6G) technologies. A key area of current investigation involves the integration of reconfigurable intelligent surface (RIS) into wireless systems to enhance spectral and energy efficiency. These efforts have led to the establishment of fundamental frameworks encompassing performance analysis, system architecture, and optimization strategies for RIS-assisted networks, providing a solid foundation for future research \cite{zhu2023performance,li2024toward,li2025transmissive,zhu2024performance,zhang2024intelligent}. In spite of this, supporting large-scale multiple access and achieving higher throughput remain central goals in 6G design. To this end, the exploration of advanced multiple access schemes and efficient channel enhancement techniques represents an essential and ongoing research trajectory.

Conventional multiple access schemes, such as time division multiple access (TDMA), differentiate users by allocating orthogonal resources, while space division multiple access (SDMA) achieves user multiplexing through multi-antenna beamforming. However, with the continuous growth in user density and data demand, conventional multiple access techniques are increasingly incapable of satisfying these escalating requirements. Non-orthogonal multiple access (NOMA) addresses this challenge by introducing the power domain as an additional multiplexing resource. It relies on two key mechanisms: superposition coding (SC) and successive interference cancellation (SIC) \cite{TCover1972Broad, PWW1998VBL, pp1994ana, mfh2016min}. In last decades, research on the mechanisms of NOMA and its performance in wireless communication systems has advanced significantly \cite{ding2016imp, ffa2017jot, yang2016gen, sun2017opt}. Recent advancements in NOMA have emphasized its integration with emerging technologies. Specifically, authors in \cite{mu2020exp, zheng2020int, li2022joint} have investigated the design of system optimization frameworks for NOMA systems incorporating RIS, and researchers in \cite{ngu2023cha, huang2025att} have explored the integration of neural networks with NOMA.

The movable antenna (MA), also referred to as the fluid antenna system, represents a novel antenna architecture characterized by its core innovation, i.e., antenna mobility. By enabling dynamic repositioning, MAs introduce spatial diversity that facilitates channel gain enhancement through optimized antenna placement. A comprehensive tutorial on recent MA research is provided in \cite{new2024tutorial}, outlining key techniques, emerging opportunities, and existing challenges. To support system-level modeling, a field-response channel model was developed in \cite{zhu2024mpd}, capturing the coupling between antenna movement and multipath-induced phase shifts, which significantly impact the base station (BS)–user channel response. The performance of MA-assisted wireless systems has been extensively analyzed from an outage probability perspective. In \cite{new2024information}, an information-theoretic framework for MA-MIMO systems is proposed, addressing system optimization, the diversity-multiplexing tradeoff, and outage characterization. To reduce computational complexity, \cite{xu2024revisiting} presents two approximation methods for MA-correlated channels. In \cite{new2024fluid}, the outage performance and diversity gain of point-to-point MA systems are examined in detail. Additionally, the deployment of MAs has also elevated the importance of efficient channel estimation techniques. In \cite{xu2024channel}, a low-complexity, high-accuracy estimation scheme is developed for multiuser MA systems. And in \cite{ma2023compressed}, the authors propose a successive transmitter–receiver compressed sensing approach, effectively reducing pilot overhead and computational burden. Another key research focus lies in joint optimization strategies of MA positioning for various communication scenarios \cite{ma2024movable, liu2025movable, zhu2024movantena, xu2025capacity, qin2024ant, zhang2025sumrate}. In \cite{ma2024movable}, the integration of MAs into sensing systems is explored through one-dimensional and two-dimensional array configurations, with an MA position optimization algorithm developed to enhance sensing accuracy—validated through numerical simulations. For uplink communication, where users are equipped with MAs and the BS employs fixed-position antennas (FPAs), \cite{zhu2024movantena} proposes a joint optimization algorithm for MA positioning and receive matrix design, aiming to minimize total transmission power while satisfying users' minimum rate constraints. Numerical simulation confirms the energy-saving benefits of MAs. Similarly, for downlink communication with the same MA-user/FPA-BS configuration, \cite{qin2024ant} introduces an algorithm that jointly optimizes BS beamforming and MA positioning to minimize transmission power, again demonstrating the effectiveness of MAs in reducing power consumption.

In NOMA-based multiuser systems, both user clustering and intra-cluster SIC ordering heavily rely on the strength of channel gain between the BS and users. To enhance system performance, researchers have investigated the integration of channel-boosting techniques into NOMA, such as RIS. More recently, the emergence of MAs has opened a promising direction for advancing NOMA systems. By enabling dynamic antenna positioning, MAs offer the flexibility to suppress inter-cluster interference and enhance channel disparities among intra-cluster users, which are critical for efficient SIC. Despite these advantages, integrating MAs into NOMA poses several technical challenges: (1) determining whether user clustering should adapt dynamically with MA position optimization or remain fixed; (2) designing joint optimization strategies for SIC ordering and MA positioning, given the location-dependent nature of channel gains; and (3) developing low-complexity algorithms capable of addressing the strong coupling among MA positions, power allocation, and beamforming in typical optimization frameworks. Preliminary investigations into MA-NOMA integration have been conducted. In \cite{zhou2024mov}, the authors investigate the role of MAs in downlink intra-cluster NOMA, considering two single-MA users served by an FPA-equipped BS. The authors formulate a sum-rate maximization problem and demonstrate that their proposed algorithm significantly enhances user rates and reduces outage probability. Similarly, in \cite{li2024sum}, the authors examine uplink intra-cluster NOMA in a scenario where an FPA-equipped BS serves multiple single-MA users. The study jointly optimizes MA positions, SIC ordering, and transmit power to maximize the uplink sum rate. Their findings revealed that MA integration significantly improves system capacity in uplink NOMA. Furthermore, \cite{he2024mov, gao2024mov} examined NOMA-MA integration in other contexts. The study in \cite{he2024mov} focused on optimizing MA positioning, power allocation, and rate allocation to enhance NOMA short-packet transmission performance. Meanwhile, \cite{gao2024mov} investigated MA positioning strategies in wireless-powered communication networks (WPCN) to optimize both downlink energy transfer and uplink NOMA data transmission.

Despite recent advances, the scenario in which both BS and users are equipped with MAs for enabling NOMA-based multiuser communications remains underexplored, which demonstrates potential for applications in wireless relay communications and wireless backhaul communications \cite{tang2015robust, wang2022task, arslan2025design}. To bridge this research gap, we propose a joint optimization algorithm to maximize the throughput of an MA-assisted downlink NOMA multi-user multiple-input-single-output (MISO) system. The objective is achieved by jointly optimizing the transmit and receive antenna position, the transmit beamforming, and power allocation. To ensure the successful execution of SIC, the optimization problem is formulated as a max-min problem. However, due to the interdependence of MA positions, beamforming vectors, and power allocation parameters, this problem is difficult to address directly. To this end, we employ the sequential parametric convex approximation (SPCA) method, which is specifically designed to solve optimization problems with multiple non-convex constraints \cite{beck2010seq}. The main contribution of this paper are summarized as follows:

\begin{itemize}
	\item[1)] We propose an MA-enabled multi-user NOMA system,  where the BS employs a MA array while each user is equipped with a single MA to enhance the performance of downlink multiuser NOMA wireless communications. Based on the field-response channel model, we characterize the downlink channel for this MA-enabled configuration at both the BS and user end (UE). Adopting a distance-based SIC decoding order, we formulate a joint optimization problem to maximize system throughput by simultaneously optimizing the BS beamforming vectors, power allocation, and the MA positions at both the BS and user ends.
	
	\item[2)] The strong coupling among optimization parameters poses significant challenges for effective problem solving. To tackle the resultant non-convex problem, we propose an algorithm based on a dual-loop framework. The outer loop, following the AO principle, mitigates the strong coupling among variables by decomposing the original problem into three subproblems, which are solved sequentially and iterated until convergence. The inner loop tackles each subproblem individually, converting non-convex constraints into equivalent or sufficient convex approximations using SCA and SCPA. The proposed algorithm ensures stable convergence to a locally optimal solution.
	
	\item[3)] We conduct extensive numerical simulations to evaluate the performance gains enabled by incorporating MAs at both the BS and user ends in downlink NOMA multiuser systems. The results demonstrate that the proposed scheme consistently outperforms SDMA, TDMA, and conventional NOMA approaches across a wide range of operational scenarios, including varying power budgets, BS antenna configurations, and user population sizes.
\end{itemize}

The remainder of this paper is organized as follows. In Section \ref{section2}, we introduce the MA-assisted NOMA system and the exact channel model. Section \ref{section3} provides the solution of the formulated problem. Section \ref{section4} then shows the simulation results. Finally, Section \ref{section5} draws some conclusion remarks.

\textbf{Notations:} Boldface upper and lower case letters denote matrices and column vectors respectively. $(\mathbf{\cdot})^T$ and $(\mathbf{\cdot})^H$ denote transpose and conjugate transpose, respectively. $\vert \mathbf{\cdot} \vert$ and $\Vert \mathbf{\cdot} \Vert_2$ denote the absolute value and Euclidean norm, respectively. $\Re(\mathbf{\cdot})$ and $\Im(\mathbf{\cdot})$ represent the real part and image part, respectively. $\mathbf{I}_N$ is the identical matrix of size $N \times N$. $\mathcal{CN}(0, \sigma^2)$ denotes the circular symmetric complex Gaussian (CSGN) distribution with mean zero and variance $\sigma^2$. $\mathcal{U}[a, b]$ represents the uniform distribution between the real-number $a$ and $b$. $\mathbb{R}$ and $\mathbb{C}$ denote the sets of real and complex numbers, respectively. $\mathbb{E}(\mathbf{\cdot})$ is the expected value of a random variable. The notation ${\rm diag}(\mathbf{\cdots})$ denotes a diagonal matrix whose diagonal entries are the elements enclosed within the parentheses.

\section{System Model and Problem Formulation} \label{section2}
\subsection{System Model}
\begin{figure}
	\centering
	\includegraphics[width = 1\linewidth]{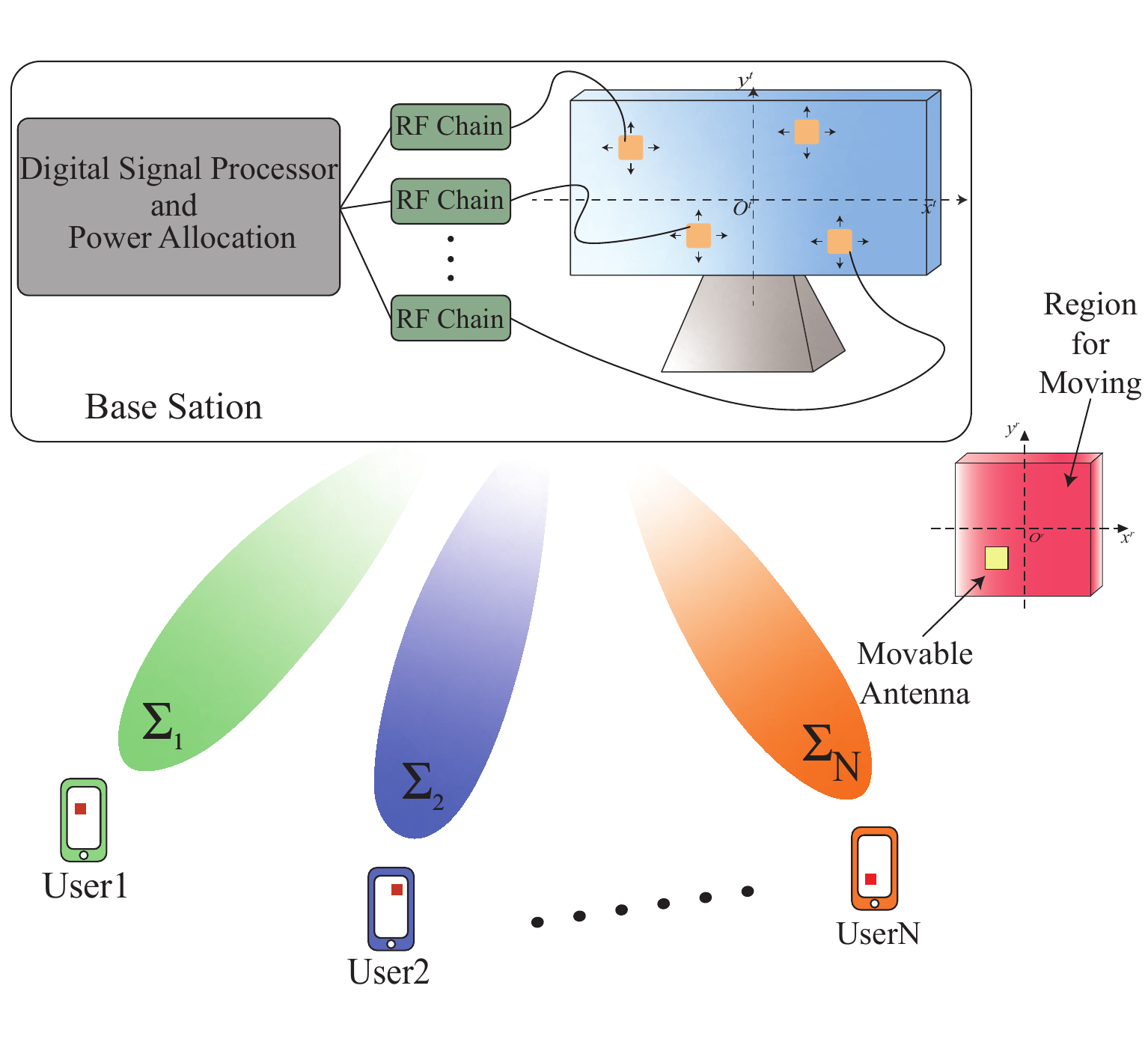}
	\caption{MA-assisted multi-user NOMA downlink MISO communication system.}
	\label{system_model}
\end{figure}

As illustrated in Fig. \ref{system_model}, this work investigates an MA-assisted multi-user NOMA downlink MISO communication system. Specifically, the BS is equipped with $M$ MAs and serves $N$ users, each with a single MA, indexed by $\mathcal{M}=\{1,\cdots, M\}, \mathcal{N}=\{1, \cdots, N\}$. The MAs at the BS are distributed over a planar area of size $R_t \times R_t$, while each user’s MA is confined within a square region of size $R_r \times R_r$. For clarity, we denote the planar area at BS as $\mathcal{C}_t$ and establish a Cartesian coordinate system centered at its midpoint. Accordingly, the position of the $m$-th MA at the BS within this area is represented as $\mathbf{u}_{t,m} = [x^t_m, y^t_m] \in \mathcal{C}_t, \mathbf{u}_{t,m} \in \mathbb{R}^{1 \times 2}, m \in \mathcal{M}$. Similarly, we define $\mathcal{C}_{r,n}$ as the movement region of the $n$-th user’s MA, with a coordinate system centered at its midpoint, allowing the MA's position to be expressed as $\mathbf{u}_{r,n}=[x_n^r,y_n^r]\in \mathcal{C}_{r,n},\mathbf{u}_{r,n} \in \mathbb{R}^{1\times 2}, n\in\mathcal{N}$. The BS transmits signals to the users via MAs, routing them through flexible cables after allocating power coefficients. Upon reception, the MAs at UEs receive the signals through flexible cables to local processors, where the desired signals are decoded. The received signal can be given by:
\begin{equation}\label{equa-1}
	\mathbf{y} = \mathbf{H}\left(\tilde{\mathbf{u}}_t, \tilde{\mathbf{u}}_r\right) \mathbf{W} \mathbf{P} \mathbf{s} + \mathbf{n},
\end{equation}
where $\mathbf{y} = [y_1,y_2,\cdots,y_N]^T \in \mathbb{C}^{N \times 1}$ is the received signal vector; $\mathbf{H}\left(\tilde{\mathbf{u}}_t, \tilde{\mathbf{u}}_r\right) = \left(\mathbf{h}^T_1(\tilde{\mathbf{u}}_t, \mathbf{u}_{r,1}), \cdots, \mathbf{h}^T_N(\tilde{\mathbf{u}}_t, \mathbf{u}_{r,N})\right)^T \in \mathbb{C}^{N\times M}$ is the channel response matrix (CRM), which is jointly determined by the MA positions $\tilde{\mathbf{u}}_t = (\mathbf{u}^T_{t,1}, \cdots, \mathbf{u}^T_{t,M})^T \in \mathbb{R}^{M \times 2}$ at the BS and $\tilde{\mathbf{u}}_r = (\mathbf{u}^T_{r_1}, \cdots, \mathbf{u}^T_{r,N}) \in \mathbb{R}^{N\times 2}$ at the users; the beamforming matrix, $\mathbf{W} = (\mathbf{w}_1, \cdots, \mathbf{w}_N) \in \mathbb{C}^{M \times N}$, satisfies the condition $\Vert \mathbf{w}_n \Vert_2^2 \leqslant 1, n \in \mathcal{N}$; moreover, $\mathbf{P} = {\rm diag}(\sqrt{p_1}, \cdots, \sqrt{p_N}) \in \mathbb{R}^{N \times N}$ is the matrix composed of power allocation coefficients, where $p_n, n\in \mathcal{N}$ denotes the coefficient assigned to the $n$-th user. $\mathbf{s} = [s_1, \cdots, s_N]^T \in \mathbb{C}^{N\times 1}$ represents the signal symbol vector sent to users with normalized power, i.e., $\mathbb{E}(\mathbf{s}\mathbf{s}^H)=\mathbf{I}_N$. $\mathbf{n} = [n_1, \cdots, n_N]^T \in \mathbb{C}^{N \times 1}$ is the addictive white Gaussian noise (AWGN) vector in received signal with average power $\sigma^2$, i.e., $\mathbf{n} \sim \mathcal{CN}\left(\mathbf{0}, \sigma^2\mathbf{I}_N\right)$. Specifically, the signal received by the MA of $n$-th user can be expressed as:
\begin{equation} \label{equa-2}
	y_n =  \mathbf{h}_n(\tilde{\mathbf{u}}_t, \mathbf{u}_{r,n}) \sum_{k=1}^{N} \sqrt{p_k} \mathbf{w}_k s_k + n_n, n \in \mathcal{N}.
\end{equation}

In the communication scenario under consideration, the dimensions of the BS plane and the region within which the user's MA can move are significantly smaller than the distance between the BS and the user. Consequently, the far-field response channel model proposed in \cite{zhu2024mpd} can be applied to characterize the BS-user channel. This implies that MA movement does not affect the angle of departure (AoD), angle of arrival (AoA), or path loss factor of the channel. However, adjusting the positions of MAs at both the BS and users alters the signal propagation distance, influencing its arrival delay and consequently modifying the received signal phase. The superposition of phase shifts from signals traveling along different paths results in variations in the channel response between the user and BS.

To facilitate the description of the impact of MAs on channel response, we define the origin of the BS plane as $\mathbf{o}_t = [0, 0]$ and the origin of the $n$-th user as $\mathbf{o}_{r,n}=[0, 0]$. The number of transmission and reception paths between $\mathbf{o}_t$ and $\mathbf{o}_{r, n}$ are denoted as $L^t_n$ and $L^r_n$ respectively. Let $\bm{\theta}^t_n = [\theta^t_{n,1}, \cdots, \theta^t_{n,L^t_n}]$ and $\bm{\phi}^t_n = [\phi^t_{n,1}, \cdots, \phi^t_{n, L^t_n}]$ represent the elevation and azimuth vectors of the AoDs for transmission paths. The path difference introduced by the $m$-th MA at the BS relative to $\mathbf{o}_t$ on the $i$-th transmission path is denoted as $\rho^t_{n, m, i} = x^t_m\cos\theta^t_{n,i}\cos\phi^t_{n,i} + y^t_m\cos\theta^t_{n, i}\sin\phi^t_{n,i}$. Consequently, the path difference vector introduced by the $m$-th MA at BS is represented as $\bm{\rho}^t_{n, m} = [\rho^t_{n,m,1}, \cdots, \rho^t_{n,m,L^t_n}]^T$. Subsequently, the transmission field response vector (FRV) of the $m$-th MA at BS is given by:
\begin{equation}\label{equa-3}
	\mathbf{g}_{n,m}(\bm{\rho}^t_{n,m}) = \left[e^{j\frac{2\pi}{\lambda_c}\rho^t_{n,m,1}}, \cdots, e^{j\frac{2\pi}{\lambda_c}\rho^t_{n,m,L^t_n}} \right]^T, m \in \mathcal{M},                                                                               
\end{equation}
where $\lambda_c$ is the wavelength of transmission carrier. Then, the field response matrix between the $n$-th user and BS can be expressed as:
\begin{equation}\label{equa-4}
	\mathbf{G}_n = \left(\mathbf{g}_{n,1}, \mathbf{g}_{n,2}, \cdots, \mathbf{g}_{n,M}\right), n \in \mathcal{N}.
\end{equation}

Similarly, let $\bm{\theta}^r_n = [\theta^r_{n,1}, \cdots, \theta^r_{n,L^r_n}]$ and $\bm{\phi}^r_n = [\phi^r,_{n,1}, \cdots, \phi^r_{n, L^r_n}]$ denote the elevation and azimuth angle vectors of AoAs corresponding to the reception paths between the $n$-th user and BS, respectively. The path difference introduced by the position of $n$-th user's MA on the $j$-th reception path, relative to $\mathbf{o}_{r,n}$, is given by $\rho^r_{n, j} = x^r_n\cos\theta^r_{n,j}\cos\phi^r_{n,j} + y^r_n\cos\theta^r_{n,j}\sin\phi^r_{n,j}$. The vector $\bm{\rho}^r_n = [\rho^r_{n,1}, \cdots, \rho^r_{n, L^r_n}]$ represents the path differences introduced by the $n$-th user's MA across all receiving paths. Consequently, the reception FRV of the $n$-th user's MA can be expressed as:
\begin{equation}\label{equa-5}
	\mathbf{f}_n(\bm{\rho}^r_n) = \left[e^{j\frac{2\pi}{\lambda_c}\rho^r_{n,1}}, e^{j\frac{2\pi}{\lambda_c}\rho^r_{n,2}}, \cdots, e^{j\frac{2\pi}{\lambda_c} \rho^r_{n,L^r_n}}\right]^T, n\in\mathcal{N}.
\end{equation}
Therefore, the channel response vector between BS and $n$-th user is given by:
\begin{equation}\label{equa-6}
	\begin{split}
		\mathbf{h}_n(\tilde{\mathbf{u}}_t, \mathbf{u}_{r,n}) & = \mathbf{f}^T_n(\bm{\rho}^r_n) \mathbf{\Sigma}_n \mathbf{G}_n \\
		 = & \left(\begin{matrix}
			\sum_{i=1}^{L^t_n}\sum_{j=1}^{L^r_n}\Sigma^{(n)}_{j,i} e^{j\frac{2\pi}{\lambda_c}(\rho^t_{n,1,i}+\rho^r_{n,j})}\\
			\mathbf{\cdots} \\
			\sum_{i=1}^{L^t_n}\sum_{j=1}^{L^r_n}\Sigma^{(n)}_{j,i} e^{j\frac{2\pi}{\lambda_c}(\rho^t_{n,M,i}+\rho^r_{n,j})}
		\end{matrix}\right)^T,
	\end{split}
\end{equation}
where $\mathbf{\Sigma}_n$ represents the path response matrix (PRM) between the origin $\mathbf{o}_t$ at BS and $\mathbf{o}_{r,n}$ of the $n$-th user. Additionally, $\Sigma^{(n)}_{j,i}$ denotes the path response coefficient (PRC) from $\mathbf{o}_t$, through the $i$-th transmission path and the $j$-th reception path, to the $\mathbf{o}_{r, n}$.

On the other hand, in NOMA systems, the order of SIC significantly impacts the upper performance limit of the system. Therefore, determining the optimal SIC order in different scenarios is a crucial research topic in NOMA technique. This paper focuses on designing a joint optimization algorithm for antenna positioning, beamforming, and power allocation under a fixed SIC order to achieve optimal system performance. Specifically, the SIC order is determined based on the distance between users and BS, with users farther from the BS assigned earlier SIC orders. In this scheme, the first user is the farthest from the BS, while the last user is the closest. During signal reception, the first user treats all signals intended for other users as interference, making its maximum achievable reception rate as
\begin{equation}\label{equa-7}
	R^1_1 = \log_2 \left(1 + \frac{\left|\sqrt{p_1} \mathbf{h}_1(\tilde{\mathbf{u}}_t, \mathbf{u}_{r, 1}) \mathbf{w}_1\right|^2}{\sum_{j=2}^{N}\left|\sqrt{p_j} \mathbf{h}_1(\tilde{\mathbf{u}}_t, \mathbf{u}_{r, 1}) \mathbf{w}_j\right|^2 + \sigma^2}\right).
\end{equation}
For the $k$-th user, signals intended for preceding users must first be decoded. The maximum achievable separation rate for the signal intended for the $l$-th $(l<k)$ user at the $k$-th user is denoted as
\begin{equation}\label{equa-8}
	R^l_k = \log_2 \left(1 + \frac{\left|\sqrt{p_l} \mathbf{h}_k(\tilde{\mathbf{u}}_t, \mathbf{u}_{r, k}) \mathbf{w}_l\right|^2}{\sum_{j=l+1}^{N}\left|\sqrt{p_j} \mathbf{h}_k(\tilde{\mathbf{u}}_t, \mathbf{u}_{r, k}) \mathbf{w}_j\right|^2 + \sigma^2}\right).
\end{equation}
Meanwhile, signals intended for subsequent users are treated as interference, and the achievable reception rate of the $k$-th user is
\begin{equation}\label{equa-9}
	R^k_k = \log_2 \left(1 + \frac{\left|\sqrt{p_k} \mathbf{h}_k(\tilde{\mathbf{u}}_t, \mathbf{u}_{r, k}) \mathbf{w}_k\right|^2}{\sum_{j=k+1}^{N}\left|\sqrt{p_j} \mathbf{h}_k(\tilde{\mathbf{u}}_t, \mathbf{u}_{r, k}) \mathbf{w}_j\right|^2 + \sigma^2}\right).
\end{equation}
To ensure successful SIC, the maximum achievable rate for the $k$-th user is given by the minimum of its own maximum achievable reception rate $R^k_k$ and the maximum achievable separation rates at subsequent users $R^k_l (l>k)$, expressed as
\begin{equation}\label{equa-10}
	R_k = \min\{R^k_k, R^k_{k+1}, \mathbf{\cdots}, R^k_N\}.
\end{equation}

Since the channel response strength is inversely proportional to the distance between the user and BS, users farther from the BS may experience near-zero data rates if no precautions are taken. To address this, we impose constraints on antenna positioning, beamforming, and power allocation, specifically
\begin{equation}\label{equa-11}
	\left|\sqrt{p_1}\mathbf{h}_k\mathbf{w}_1\right|^2 \geqslant \alpha \left|\sqrt{p_2}\mathbf{h}_k\mathbf{w}_2\right|^2 \geqslant \mathbf{\cdots} \geqslant \alpha^{k-1}\left|\sqrt{p_k}\mathbf{h}_k\mathbf{w}_k\right|^2,
\end{equation}
where, $k \in \mathcal{N}$ and $\alpha (\alpha \geqslant 1)$ is a coefficient that controls the minimum rate threshold (MRT) for users far from the BS. An increase in $\alpha$ corresponds to a higher minimum rate for these users. The constraint (\ref{equa-11}) arises because users farther from the BS typically exhibit weaker channel response strengths, i.e., $\Vert \mathbf{h}_1 \Vert_2^2 \leqslant \Vert \mathbf{h}_2 \Vert_2^2 \leqslant \mathbf{\cdots} \leqslant \Vert \mathbf{h}_N \Vert_2^2$. Without constraint (\ref{equa-11}), system throughput would likely be maximized by allocating nearly all communication resources to users with higher channel strength and lower interference. Consider a system with three users, numbered according to the SIC order, with corresponding signal-to-interference-plus-noise ratios (SINRs) of
\begin{equation}\label{equa-12}
	\begin{split}
		& \quad \gamma^1_1 = \frac{\left|\sqrt{p_1}\mathbf{h}_1\mathbf{w}_1\right|^2}{\left|\sqrt{p_2}\mathbf{h}_1\mathbf{w}_2\right|^2 + \left|\sqrt{p_3}\mathbf{h}_1\mathbf{w}_3\right|^2 + \sigma^2}, \\
		&\begin{cases}
			\gamma^1_2 = \dfrac{\left|\sqrt{p_1}\mathbf{h}_2\mathbf{w}_1\right|^2}{\left|\sqrt{p_2}\mathbf{h}_2\mathbf{w}_2\right|^2 + \left|\sqrt{p_3}\mathbf{h}_2\mathbf{w}_3\right|^2 + \sigma^2},\\
			\gamma^2_2 = \dfrac{\left|\sqrt{p_2}\mathbf{h}_2\mathbf{w}_2\right|^2}{\left|\sqrt{p_3}\mathbf{h}_2\mathbf{w}_3\right|^2 + \sigma^2},
		\end{cases}\\
		&\begin{cases}
			\gamma^1_3 = \dfrac{\left|\sqrt{p_1}\mathbf{h}_3\mathbf{w}_1\right|^2}{\left|\sqrt{p_2}\mathbf{h}_3\mathbf{w}_2\right|^2 + \left|\sqrt{p_3}\mathbf{h}_3\mathbf{w}_3\right|^2 + \sigma^2}, \\
			\gamma^2_3 = \dfrac{\left|\sqrt{p_2}\mathbf{h}_3\mathbf{w}_2\right|^2}{\left|\sqrt{p_3}\mathbf{h}_3\mathbf{w}_3\right|^2 + \sigma^2}, \\
			\gamma^3_3 = \dfrac{\left|\sqrt{p_3}\mathbf{h}_3\mathbf{w}_3\right|^2}{\sigma^2}.
		\end{cases}
	\end{split}
\end{equation}
The maximum achievable rate for the first user is determined by $\min\{\gamma^1_1, \gamma^1_2, \gamma^1_3\}$. By introducing the constraint (\ref{equa-11}), we ensure that the numerators of $\gamma^1_2$ and $\gamma^1_3$ are larger than the interference terms in their respective denominators, thus establishing a minimum threshold for $\gamma^1_2$ and $\gamma^1_3$. This helps ensure that the first user can be allocated an effective data rate. Therefore, (\ref{equa-11}) effectively mitigates the issue of disproportionate resource allocation to users with high channel strength and low interference.

\subsection{Problem Formulation}

In this paper, we aim to maximize the throughput of an MA-assisted multi-user downlink NOMA system by optimizing the positions of the MAs at both the BS and user ends, as well as beamforming and power allocation. The throughput for each user is determined by the minimum of its achievable reception rate and its separation rate. While maximizing overall system throughput, it is essential to ensure that the data requirements of all users are satisfied, enabling efficient resource allocation and guaranteeing a minimum throughput for each user. Consequently, we formulate the optimization problem as (P0).
\begin{subequations}
	\begin{align}
		\text{(P0)}& \qquad \max_{\tilde{\mathbf{u}}_t, \tilde{\mathbf{u}}_r, \mathbf{W}, \mathbf{P}} \quad \sum_{k = 1}^{N} R_k, \label{sub0-1}\\
		\text{s.t.}&  \enspace \left|\sqrt{p_1}\mathbf{h}_n\mathbf{w}_1\right|^2 \geqslant \mathbf{\cdots} \geqslant \alpha^{n-1}\left|\sqrt{p_n}\mathbf{h}_n\mathbf{w}_n\right|^2,n\in \mathcal{N}, \label{sub0-2}\\
		& \qquad\qquad\quad \Vert \mathbf{w}_n \Vert_2 \leqslant 1, n \in \mathcal{N}, \label{sub0-3}\\
		& \qquad\qquad\qquad \sum_{n=1}^{N}p_n \leqslant P_t,\label{sub0-4}\\
		& \qquad\qquad\quad \mathbf{u}_{t,m} \in \mathcal{C}_t, m\in \mathcal{M}, \label{sub0-5}\\ 
		& \qquad  \Vert \mathbf{u}_{t, i} - \mathbf{u}_{t, l}\Vert_2 \geqslant \frac{1}{2}\lambda_c, i, l\in \mathcal{M}, i \neq l, \label{sub0-6}\\
		& \qquad\qquad\quad \mathbf{u}_{r, n}  \in \mathcal{C}_{r, n}, n\in \mathcal{N}, \label{sub0-7}
	\end{align}
\end{subequations}
where (\ref{sub0-4}) represents the constraint on total transmit power at BS, while (\ref{sub0-5}) and (\ref{sub0-6}) impose constraints on the positions of MAs at the BS. Considering their physical size, the distance between any two MAs must exceed a specified threshold. (\ref{sub0-7}) ensures that the MAs at users are within a designated region. Since the optimization variables are coupled in (\ref{sub0-1}), (\ref{sub0-2}) which are simultaneously non-convex, solving the problem directly is challenging. Therefore, we apply sequential convex programming (SCP), SCA, and AO to decompose the problem into a series of tractable subproblems, ultimately leading to a stable solution.

\section{Proposed Solution} \label{section3}
\subsection{Problem Reformulation}
We note that a new variable $\mathbf{V} = (\mathbf{v}_1, \mathbf{v}_2, \cdots, \mathbf{v}_N) \in \mathbb{C}^{M\times N}$ can be introduced to replace the existing optimization variables $\mathbf{W}$ and $\mathbf{P}$, where their relationship is defined as $\mathbf{v}_n = \sqrt{p_n}\mathbf{w}_n, n\in\mathcal{N}$, subject to the constraints $\sum_{n=1}^{N}p_n \leqslant P_t$ and $\Vert \mathbf{w}_n \Vert_2 \leqslant 1, n\in\mathcal{N}$. Consequently, $\mathbf{V}$ must satisfy $\sum_{n=1}^{N}\Vert \mathbf{v}_n \Vert_2^2 \leqslant P_t$. Once the optimized solution for $\hat{\mathbf{V}}$ is obtained, the corresponding values of $\hat{\mathbf{P}}$ and $\hat{\mathbf{W}}$ can be derived as $\hat{p}_n = \Vert \hat{\mathbf{v}}_n \Vert_2^2$ and $ \hat{\mathbf{w}}_n = \frac{\hat{\mathbf{v}}_n}{\Vert \hat{\mathbf{v}}_n \Vert_2}, n\in \mathcal{N}$, ensuring that the constraints (\ref{sub0-3}) and (\ref{sub0-4}) are met. Furthermore, we note that the objective function $\sum_{k=1}^{N}R_k = \sum_{k=1}^{N}\log_2(1+min\{\gamma^k_k, \gamma^k_{k+1}, \cdots, \gamma^k_N\})$ in problem (P0) is non-concave with respect to the optimization variables $\tilde{\mathbf{u}}_t, \tilde{\mathbf{u}}_r, \mathbf{W}$ and $\mathbf{P}$. To address this issue, we introduce auxiliary variables $\mathbf{r} \in \mathcal{R}^{N}$ and $\bm{\nu} \in \mathcal{R}^{N\times N}$, where $r_n - 1 \leqslant min\{\gamma^n_n, \gamma^n_{n+1}, \cdots, \gamma^n_N\}$ and $\nu_{k,n}$ is larger than the denominator of $\gamma^n_k$, i.e.,
\begin{equation}\label{equa-14}
	\begin{split}
		 r_n - 1 \leqslant \frac{\left|\sqrt{p_n}\mathbf{h}_k\mathbf{w}_n\right|^2}{\sum_{j=n+1}^{N}\left|\sqrt{p_j}\mathbf{h}_k\mathbf{w}_j\right|^2 + \sigma^2}, \begin{cases}
		 	n\in\{1,\cdots,N-1\},\\
		 	k\in\{n,\cdots,N\},
		 \end{cases}
	\end{split}
\end{equation}
\begin{equation}\label{equa-15}
	\nu_{k,n} \geqslant \sum_{j=n+1}^{N}\left|\sqrt{p_j}\mathbf{h}_k\mathbf{w}_j\right|^2+\sigma^2,\begin{cases}
		n \in \{1,\cdots,N-1\},\\
		k \in \{n,\cdots, N\},
	\end{cases}
\end{equation}
\begin{equation}\label{equa-16}
	r_n - 1 \leqslant \frac{\left|\sqrt{p_n}\mathbf{h}_k\mathbf{w}_n\right|^2}{\sigma^2},n=k=N,
\end{equation}
\begin{equation}\label{equa-17}
	\nu_{k,n} \geqslant \sigma^2, n=k=N.
\end{equation}
Therefore, we can transform (P0) into a more tractable form, as presented in (P1):
\begin{subequations}
	\begin{align}
		\text{(P1)}& \qquad \qquad \max_{\tilde{\mathbf{u}}_t, \tilde{\mathbf{u}}_r, \mathbf{V}, \mathbf{r}, \bm{\nu}}\left(\prod_{k=1}^{N}r_k\right)^{\frac{1}{N}},\label{sub1-1}\\
		\text{s.t.}& \begin{cases}
			(r_n-1)\nu_{k,n} \leqslant \left|\mathbf{h}_k\mathbf{v}_n\right|^2,\enspace 1\leqslant n\leqslant N-1,\\
			\nu_{k,n} \geqslant \sum_{j=n+1}^{N}\left|\mathbf{h}_k\mathbf{v}_j\right|^2+\sigma^2,\enspace n\leqslant k\leqslant N,
		\end{cases}\label{sub1-2}\\
		&\qquad \begin{cases}
			(r_n-1)\nu_{k,n} \leqslant \left|\mathbf{h}_k\mathbf{v}_n\right|^2, \quad n=N\\
			\nu_{k,n} \geqslant \sigma^2, \quad k=n=N
		\end{cases}\label{sub1-3}\\
		&\quad\enspace \left|\mathbf{h}_n\mathbf{v}_1\right|^2 \geqslant \cdots \geqslant \alpha^{n-1}\left|\mathbf{h}_n\mathbf{v}_n\right|^2, n\in\mathcal{N},\label{sub1-4}\\
		& \qquad\qquad\qquad \Vert \mathbf{V} \Vert_F \leqslant \sqrt{P_t},\label{sub1-5}\\
		& \qquad\qquad\quad (\ref{sub0-5}),(\ref{sub0-6}),(\ref{sub0-7}).\notag
	\end{align}
\end{subequations}

\subsection{Beamforming and Power Allocation Optimization}
Given that the variables $\tilde{\mathbf{u}}_t, \tilde{\mathbf{u}}_r$, and $\mathbf{V}$ are coupled in (\ref{sub1-2}), (\ref{sub1-3}), and (\ref{sub1-4}), simultaneously optimizing all three variables presents significant challenges. To facilitate the attainment of a stable solution, we employ the AO algorithm, which allows for the sequential optimization of each variable while keeping the others fixed. Specifically, we first consider the positions of MAs at both the BS and user ends as fixed, and subsequently optimize the beamforming and power allocation at the BS, i.e., maintaining $\tilde{\mathbf{u}}_t$ and $\tilde{\mathbf{u}}_r$ constant and optimizing $\mathbf{V}$. Thus, we can obtain the original first sub-problem as
\begin{subequations}
	\begin{align}
		\text{(P2.0)}& \qquad \max_{\mathbf{V}, \mathbf{r}, \bm{\nu}} \left(\prod_{k=1}^{N}r_k\right)^{\frac{1}{N}}, \label{sub2.0-1}\\
		\text{s.t.}& \qquad (\ref{sub1-2}), (\ref{sub1-3}), (\ref{sub1-4}), (\ref{sub1-5}).\notag
	\end{align}
\end{subequations}
(P2.0) remains intractable due to the non-convex nature of (\ref{sub1-2}), (\ref{sub1-3}), and (\ref{sub1-4}). We reformulate these constraints into equivalent or sufficient convex forms. Leveraging concepts from SCP and SCA, we decompose the functions in (\ref{sub1-2}), (\ref{sub1-3}), and (\ref{sub1-4}) and construct their global convex upper or concave lower bounds, thereby enabling to approximate the original non-convex constraints with convex counterparts.

For (\ref{sub1-2}), we observe that the first inequality constraint is non-convex, and the left-hand side of this constraint can be expressed as: 
\begin{equation}\label{equa-20}
	\left(r_n - 1\right)\nu_{k, n} = \frac{1}{4}\left(r_n + \nu_{k, n}\right)^2 - \frac{1}{4}\left(r_n - \nu_{k, n}\right)^2 - \nu_{k, n},
\end{equation}
where $-\frac{1}{4}\left(r_n - \nu_{k, n}\right)^2$ is concave with respect to $r_n$ and $\nu_{k, n}$. Due to the concave nature, a global convex upper bound for $-\frac{1}{4}\left(r_n - \nu_{k, n}\right)^2$ an be obtained by constructing a tangent at any arbitrary point $(r_n^j,\nu_{k, n}^j)$. Consequently, the global convex upper bound for $(r_n-1)\nu_{k, n}$ can be expressed in the following form
\begin{equation}\label{equa-21}
	\begin{split}
		h(r_n, \nu_{k, n}, r_n^j, \nu_{k, n}^j) & \triangleq  \frac{1}{4}\left(r_n + \nu_{k, n}\right)^2 - \nu_{k, n} - \frac{1}{4}[(r_n^j - \nu_{k, n}^j)^2 \\ 
		& + 2(r_n^j-\nu_{k, n}^j)(r_n-r_n^j-\nu_{k, n}+\nu_{k, n}^j)].
	\end{split}
\end{equation}
Alternatively, the right-hand side of this inequality constraint can be expanded as
\begin{equation}\label{equa-22}
	\begin{split}
		\left|\mathbf{h}_k\mathbf{v}_n\right|^2 = & (\Re(\mathbf{h}_k)\Re(\mathbf{v}_n) - \Im(\mathbf{h}_k)\Im(\mathbf{v}_n))^2 \\
		&+ (\Im(\mathbf{h}_k)\Re(\mathbf{v}_n) + \Re(\mathbf{h}_k)\Im(\mathbf{v}_n))^2.
	\end{split}
\end{equation}
To offer a more concise description of the first sub-problem, we introduce parameters $\mathbf{A}_k \in \mathbb{R}^{2M\times 2M}, k\in \mathcal{N}$ and auxiliary variables $\mathbf{B}=(\bm{\beta}_1, \mathbf{\cdots}, \bm{\beta}_N)\in \mathbb{R}^{2M\times N}$, which are defined as follows:
\begin{equation}\label{equa-23}
	\begin{split}
		&\mathbf{A}_k \triangleq \left(\begin{matrix}
			\Re(\mathbf{h}_k) \enspace -\Im(\mathbf{h}_k)\\
			\Im(\mathbf{h}_k) \quad\enspace \Re(\mathbf{h}_k)
		\end{matrix}\right)^T \left(\begin{matrix}
		\Re(\mathbf{h}_k) \enspace -\Im(\mathbf{h}_k)\\
		\Im(\mathbf{h}_k) \quad\enspace \Re(\mathbf{h}_k)
		\end{matrix}\right),\\
		&\qquad\qquad\qquad\qquad \bm{\beta}_n \triangleq \left(\begin{matrix}
			\Re(\mathbf{v}_n)\\
			\Im(\mathbf{v}_n)
		\end{matrix}\right).
	\end{split}
\end{equation}
Therefore, $\left|\mathbf{h}_k\mathbf{v}_n\right|^2$ can be reformulated as $\left|\mathbf{h}_k\mathbf{v}_n\right|^2 = \bm{\beta}_n^T\mathbf{A}_k\bm{\beta}_n$. We notice that $\bm{\beta}_n^T\mathbf{A}_k\bm{\beta}_n$ is convex respect to $\bm{\beta}_n$. Owing to the nature of its convexity, a global concave lower bound for $\left|\mathbf{h}_k\mathbf{v}_n\right|^2$ can be constructed as
\begin{equation}\label{equa-24}
	g_k(\bm{\beta}_n, \bm{\beta}_n^j) \triangleq {\bm{\beta}_n^j}^T\mathbf{A}_k\bm{\beta}_n^j + 2{\bm{\beta}_n^j}^T\mathbf{A}_k^T(\bm{\beta}_n-\bm{\beta}_n^j).
\end{equation}
where $\bm{\beta}_n^j$ is an arbitrary point for $\bm{\beta}_n$.

As for the second inequality constraint in (\ref{sub1-2}), it is convex and we just need to alter its form. First, we define that $\bm{\alpha}_{k, 1}^T = (\Re(\mathbf{h}_k), -\Im(\mathbf{h}_k))\in\mathbb{R}^{2M}$ and $\bm{\alpha}_{k, 2}^T = (\Im(\mathbf{h}_k), \Re(\mathbf{h}_k))\in\mathbb{R}^{2M}$. Thus the second inequality constraint is equivalent to 
\begin{equation}\label{equa-25}
	\frac{1+\nu_{k,n}}{2}\geqslant\left\Vert \begin{matrix}
		\bm{\alpha}_{k,1}\bm{\beta}_{n+1}\\
		\mathbf{\cdots}\\
		\bm{\alpha}_{k,1}\bm{\beta}_{N}\\
		\bm{\alpha}_{k,2}\bm{\beta}_{n+1}\\
		\mathbf{\cdots}\\
		\bm{\alpha}_{k,2}\bm{\beta}_{N}\\
		1\\
		\dfrac{1-\nu_{k, n}}{2}
	\end{matrix} \right\Vert_2.
\end{equation}

The transformation applied to (\ref{sub1-3}) is similar to that used for (\ref{sub1-2}). Regarding (\ref{sub1-4}), we observed that the sequential inequalities can first be decomposed into (\ref{equa-26}). Each resulting inequality can then be uniformly transformed—by expanding and rearranging the involved functions—to arrive at (\ref{equa-27}),
\begin{equation}\label{equa-26}
	\left|\mathbf{h}_k\mathbf{v}_n\right|^2 \geqslant \alpha\left|\mathbf{h}_k\mathbf{v}_{n+1}\right|^2,k\in\mathcal{N},1\leqslant n\leqslant k-1,
\end{equation}
\begin{equation}\label{equa-27}
	\begin{split}
		m_k&(\bm{\beta}_n,\bm{\beta}_{n+1}) \triangleq \alpha\bm{\beta}_{n+1}^T\mathbf{A}_k\bm{\beta}_{n+1}-\bm{\beta}_n^T\mathbf{A}_k\bm{\beta}_n\\
		&=\left(\begin{matrix}
			\bm{\beta}_n\\
			\bm{\beta}_{n+1}
		\end{matrix}\right)^T 
		\left(\begin{matrix}
		-\mathbf{A}_k \quad \mathbf{0}\enspace\\
		\quad\mathbf{0} \quad \alpha\mathbf{A}_k
		\end{matrix}\right) 
		\left(\begin{matrix} 
		\bm{\beta}_n\\
		\bm{\beta}_{n+1}
		\end{matrix}\right)\leqslant 0.
	\end{split}
\end{equation}
Compute the first and second derivatives of $m_k(\bm{\beta}_n,\bm{\beta}_{n+1})$ with respect to $\bm{\beta}_n$ and $\bm{\beta}_{n+1}$, thereby obtaining the gradient and Hessian matrix of $m_k(\bm{\beta}_n,\bm{\beta}_{n+1})$, as expressed below:
\begin{equation}\label{equa-28}
	\begin{split}
		\triangledown m_k(\bm{\beta}_n,\bm{\beta}_{n+1}) = 2 \left(\begin{matrix}
			-\mathbf{A}_k \quad \mathbf{0}\enspace\\
			\quad\mathbf{0} \quad \alpha\mathbf{A}_k
		\end{matrix}\right) 
		\left(\begin{matrix} 
			\bm{\beta}_n\\
			\bm{\beta}_{n+1}
		\end{matrix}\right),\\
		\triangledown^2  m_k(\bm{\beta}_n,\bm{\beta}_{n+1}) = 2 \left(\begin{matrix}
			-\mathbf{A}_k \quad \mathbf{0}\enspace\\
			\quad\mathbf{0} \quad \alpha\mathbf{A}_k
		\end{matrix}\right).\qquad
	\end{split}
\end{equation}
Let $\lambda_k$ denote the largest eigenvalue of the Hessian matrix $\triangledown^2  m_k(\bm{\beta}_n,\bm{\beta}_{n+1})$. By Taylor's theorem, given any arbitrary point $(\bm{\beta}_n^j,\bm{\beta}_{n+1}^j)$,  a global convex upper bound for $m_k(\bm{\beta}_n,\bm{\beta}_{n+1})$ can be constructed as
\begin{equation}\label{equa-29}
	\begin{split}
		f_k(&\bm{\beta}_n,\bm{\beta}_{n+1},\bm{\beta}_n^j,\bm{\beta}_{n+1}^j) \triangleq 
		\left(\begin{matrix}
			\bm{\beta}_{n}^j\\
			\bm{\beta}_{n+1}^j
		\end{matrix}\right)^T
		\left(\begin{matrix}
			-\mathbf{A}_k \enspace \mathbf{0}\enspace\\
			\enspace\mathbf{0} \quad \alpha\mathbf{A}_k
		\end{matrix}\right)
		\left(\begin{matrix}
			\bm{\beta}_{n}^j\\
			\bm{\beta}_{n+1}^j
		\end{matrix}\right)\\
		&+ \left(\begin{matrix}
			\bm{\beta}_{n}^j\\
			\bm{\beta}_{n+1}^j
		\end{matrix}\right)^T
		\left(\begin{matrix}
			-\mathbf{A}_k \enspace \mathbf{0}\enspace\\
			\enspace\mathbf{0} \quad \alpha\mathbf{A}_k
		\end{matrix}\right)^T
		\left[\left(\begin{matrix}
			\bm{\beta}_{n}\\
			\bm{\beta}_{n+1}
		\end{matrix}\right)-
		\left(\begin{matrix}
			\bm{\beta}_{n}^j\\
			\bm{\beta}_{n+1}^j
		\end{matrix}\right)\right]\\
		& + \frac{1}{2}\lambda_k\left[\left(\begin{matrix}
			\bm{\beta}_{n}\\
			\bm{\beta}_{n+1}
		\end{matrix}\right)-
		\left(\begin{matrix}
			\bm{\beta}_{n}^j\\
			\bm{\beta}_{n+1}^j
		\end{matrix}\right)\right]^T
		\left[\left(\begin{matrix}
			\bm{\beta}_{n}\\
			\bm{\beta}_{n+1}
		\end{matrix}\right)-
		\left(\begin{matrix}
			\bm{\beta}_{n}^j\\
			\bm{\beta}_{n+1}^j
		\end{matrix}\right)\right].
	\end{split}
\end{equation}

Upon transforming all constraints, we obtain an approximate convex formulation of the first subproblem as (P2).
\begin{subequations}
	\begin{align}
		&\text{(P2)} \qquad \max_{\mathbf{B},\mathbf{r}, \bm{\nu}}\left(\prod_{k=1}^{N}r_k\right)^{\frac{1}{N}},\label{sub2-1}\\
		&\text{s.t.} \quad \begin{cases}
			h(r_n, \nu_{k, n}, r_n^j, \nu_{k, n}^j) \leqslant g_k(\bm{\beta}_n, \bm{\beta}_n^j),\\
			(\ref{equa-25}), n\in\{1,\cdots,N-1\}, k\in\{n,\cdots,N\},
		\end{cases}\label{sub2-2}\\
		& \qquad\enspace \begin{cases}
			h(r_N, \nu_{N, N}, r_N^j, \nu_{N, N}^j) \leqslant g_N(\bm{\beta}_N, \bm{\beta}_N^j),\\
			\nu_{N, N} \geqslant \sigma^2,
		\end{cases}\label{sub2-3}\\
		& f_k(\bm{\beta}_n,\bm{\beta}_{n+1},\bm{\beta}_n^j,\bm{\beta}_{n+1}^j) \leqslant 0, \begin{cases}
			k\in \{2, \cdots, N\},\\
			1\leqslant n\leqslant k-1,
		\end{cases}\label{sub2-4}\\
		& \qquad\qquad\qquad\qquad\qquad (\ref{sub1-5}). \notag
	\end{align}
\end{subequations}

\subsection{User-end MA Position Optimization}
Subsequently, by fixing the beamforming vectors, power allocation parameters, and BS-end MA positions, we optimize the user-end MA positions to derive the initial formulation of the second sub-problem.
\begin{subequations}
	\begin{align}
		\text{(P3.0)}& \qquad \max_{\tilde{\mathbf{u}}_r, \mathbf{r}, \bm{\nu}} \left(\prod_{k=1}^{N}r_k\right)^\frac{1}{N}, \label{sub3.0-1}\\
		\text{s.t.}& \quad (\ref{sub1-2}), (\ref{sub1-3}), (\ref{sub1-4}), (\ref{sub0-7}).\notag
	\end{align}
\end{subequations}
(\ref{sub1-2}), (\ref{sub1-3}), and (\ref{sub1-4}) contain non-convex terms involving $\tilde{\mathbf{u}}_r$; therefore, these terms must be appropriately transformed. Specifically, for the left-hand side of the first inequality constraint in (\ref{sub1-2}), we adopt the same methodology as in the previous subsection to derive its global convex upper bound, as presented in (\ref{equa-24}). For the right-hand side of this inequality, we expand the expression, resulting in the following form:
\begin{equation}\label{equa-32}
	\begin{split}
		-\left|\mathbf{h}_k\mathbf{v}_n\right|^2 =& -\mathbf{f}_k^T\mathbf{\Sigma}_k\mathbf{G}_k\mathbf{v}_n\mathbf{v}_n^H\mathbf{G}_k^H\mathbf{\Sigma}_k^H\mathbf{f}_k^{*}
		\\
		=&\mathbf{f}_k^T(-\mathbf{C}_{k,n})\mathbf{f}_k^{*},
	\end{split}
\end{equation}
where $\mathbf{C}_{k,n} \triangleq \mathbf{\Sigma}_k\mathbf{G}_k\mathbf{v}_n\mathbf{v}_n^H\mathbf{G}_k^H\mathbf{\Sigma}_k^H $. Then, with respect to the second inequality constraint, we focus exclusively on the right-hand side, as the left-hand side is linear. Similarly, we begin by expanding the right-hand side.
\begin{equation}\label{equa-33}
	\begin{split}
		\sum_{j=n+1}^{N}\left|\mathbf{h}_k\mathbf{v}_j\right|^2 +& \sigma^2 = \sum_{j=n+1}^{N}\mathbf{f}_k^T\mathbf{\Sigma}_k\mathbf{G}_k\mathbf{v}_j\mathbf{v}_j^H\mathbf{G}_k^H\mathbf{\Sigma}_k^H\mathbf{f}_k^{*} + \sigma^2\\
		& = \mathbf{f}_k^T\mathbf{D}_{k, n}\mathbf{f}_k^{*} + \sigma^2,
	\end{split}
\end{equation}
where $\mathbf{D}_{k, n} \triangleq \mathbf{\Sigma}_k\mathbf{G}_k \left(\sum_{j=n+1}^{N}\mathbf{v}_j\mathbf{v}_j^H\right)\mathbf{G}_k^H\mathbf{\Sigma}_k^H $. 

The processing of (\ref{sub1-3}) follows the methodology applied to (\ref{sub1-2}). As for (\ref{sub1-4}), it is first decomposed into the formulation provided in (27), and the resulting terms are subsequently rearranged and expanded to yield the following expression:
\begin{equation}\label{equa-34}
	\begin{split}
		\alpha\left|\mathbf{h}_k\mathbf{v}_{n+1}\right|^2 &- \left|\mathbf{h}_k\mathbf{v}_{n}\right|^2 = \mathbf{f}_k^T\mathbf{\Sigma}_k\mathbf{G}_k\left(\alpha\mathbf{v}_{n+1}\mathbf{v}_{n+1}^H - \mathbf{v}_n\mathbf{v}_n^H\right) \\
		&\mathbf{G}_k^H\mathbf{\Sigma}_k^H\mathbf{f}_k^{*} = \mathbf{f}_k^T \mathbf{E}_{k, n} \mathbf{f}_k^{*},
	\end{split}
\end{equation}
where $\mathbf{E}_{k, n} \triangleq \mathbf{\Sigma}_k\mathbf{G}_k\left(\alpha\mathbf{v}_{n+1}\mathbf{v}_{n+1}^H - \mathbf{v}_n\mathbf{v}_n^H\right)\mathbf{G}_k^H\mathbf{\Sigma}_k^H $.

It can be observed that the three expanded expressions exhibit a common structural component. Accordingly, this structure is extracted to formulate an auxiliary function as follows:
\begin{equation}\label{equa-35}
	\begin{split}
		&P(\mathbf{u}_{r,k}, \mathbf{Q}) \triangleq \mathbf{f}_k^T(\mathbf{u}_{r,k})\mathbf{Q} \mathbf{f}_k^*(\mathbf{u}_{r,k})\\
		&= \sum_{i=1}^{L_r}\left|q_{i,i}\right| + 2\sum_{i=1}^{L_r-1}\sum_{j=i+1}^{L_r}\left|q_{i,j}\right|\cos[\frac{2\pi}{\lambda_c}(\rho_{k,i}^r-\rho_{k,j}^r) + \angle q_{i,j}],
	\end{split}
\end{equation}
where $q_{i,j}$ denotes the $(i,j)$-th element of matrix $\mathbf{Q}$. Global convex upper bounds for (\ref{equa-32}), (\ref{equa-33}), and (\ref{equa-34}) can be obtained by constructing a corresponding bound for $P(\mathbf{u}_{r,k}, \mathbf{Q})$. Based on Taylor's theorem, at any given point $\mathbf{u}_{r,k}^j$, the global convex upper bound of $P(\mathbf{u}_{r,k}, \mathbf{Q})$ can be expressed as
\begin{equation}\label{equa-36}
	\begin{split}
		p(\mathbf{u}_{r,k}, \mathbf{Q}, \mathbf{u}_{r,k}^j) \triangleq & P(\mathbf{u}_{r,k}^j,\mathbf{Q}) + \triangledown P(\mathbf{u}_{r,k}^j,\mathbf{Q})^T(\mathbf{u}_{r,k}-\mathbf{u}_{r,k}^j)^T\\
		+ & \frac{1}{2}\delta_{r,k}^{(\mathbf{Q})}(\mathbf{u}_{r,k}-\mathbf{u}_{r,k}^j)(\mathbf{u}_{r,k}-\mathbf{u}_{r,k}^j)^T,
	\end{split}
\end{equation}
where $\delta_{r,k}^{(\mathbf{Q})}$ satisfies $\delta_{r,k}^{(\mathbf{Q})}\mathbf{I_{L_r}} \succeq \triangledown^2 P(\mathbf{u}_{r,k}, \mathbf{Q})$. The second-order derivative of $P(\mathbf{u}_{r,k}, \mathbf{Q})$ is calculated, and upper bounds of the corresponding partial derivatives are obtained as (\ref{equa-37}), (\ref{equa-38}), and (\ref{equa-39}). By utilizing $\Vert \triangledown^2 P(\mathbf{u}_{r,k}, \mathbf{Q}) \Vert_F \geqslant \Vert \triangledown^2 P(\mathbf{u}_{r,k}, \mathbf{Q}) \Vert_2$, we derive the admissible $\delta_{r,k}^{(\mathbf{Q})}$ satisfying the condition as:
\begin{figure*}[!t]
	\begin{align}
			&\frac{\partial^2 P(\mathbf{u}_{r,k}, \mathbf{Q})}{(\partial x_k^r)^2} \leqslant \frac{8\pi^2}{\lambda_c^2}\sum_{i=1}^{L_r-1}\sum_{j=i+1}^{L_r}\left|q_{i,j}\right|\left(\cos\theta_{k,i}^r\cos\phi_{k,i}^r-\cos\theta_{k,j}^r\cos\phi_{k,j}^r\right)^2 \triangleq\zeta_{1,r,k}^{(\mathbf{Q})}, \label{equa-37}  \\
			&\frac{\partial^2 P(\mathbf{u}_{r, k}, \mathbf{Q})}{(\partial y_k^r)^2}  \leqslant \frac{8\pi^2}{\lambda_c^2}\sum_{i=1}^{L_r-1}\sum_{j=i+1}^{L_r}\left|q_{i,j}\right|\left(\cos\theta_{k,i}^r\sin\phi_{k,i}^r-\cos\theta_{k,j}^r\sin\phi_{k,j}^r\right)^2 \triangleq \zeta_{2,r,k}^{(\mathbf{Q})}, \label{equa-38} \\
			&\frac{\partial^2 P(\mathbf{u}_{r,k}, \mathbf{Q})}{\partial x_k^r \partial y_k^r} \leqslant \frac{8\pi^2}{\lambda_c^2}\sum_{i=1}^{L_r-1}\sum_{j=i+1}^{L_r}\left|q_{i,j}\right| |\cos\theta_{k,i}^r\cos\phi_{k,i}^r-\cos\theta_{k,j}^r\cos\phi_{k,j}^r| |\cos\theta_{k,i}^r\sin\phi_{k,i}^r -\cos\theta_{k,j}^r\sin\phi_{k,j}^r|\triangleq \zeta_{3,r,k}^{(\mathbf{Q})}. \label{equa-39}
	\end{align}
	\hrulefill
\end{figure*}
\begin{equation}
	\delta_{r,k}^{(\mathbf{Q})} = \left({\zeta_{1,r,k}^{(\mathbf{Q})}}^2 + {\zeta_{2,r,k}^{(\mathbf{Q})}}^2 + 2{\zeta_{3,r,k}^{(\mathbf{Q})}}^2\right)^{\frac{1}{2}}.
\end{equation}

Consequently, we arrive at the approximate convex formulation of the second subproblem, denoted as (P3).

\begin{subequations}
	\begin{align}
		\text{(P3)}& \qquad\qquad\qquad\quad \max_{\tilde{\mathbf{u}}_r, \mathbf{r}, \bm{\nu}}\left(\prod_{k=1}^{N}r_k\right)^{\frac{1}{N}}, \label{sub3-1}\\
		\text{s.t.}&  \begin{cases}
			h(r_n, \nu_{k, n}, r_n^j, \nu_{k, n}^j) + p(\mathbf{u}_{r, k}, -\mathbf{C}_{k, n}, \mathbf{u}_{r, k}^j) \leqslant 0, \\
			p(\mathbf{u}_{r, k}, \mathbf{D}_{k, n}, \mathbf{u}_{r, k}^j) + \sigma^2 - \nu_{k, n} \leqslant 0, \\
			n \in \{1, \cdots, N-1\}, k \in \{n, \cdots, N\},
		\end{cases} \label{sub3-2} \\
		& \begin{cases}
			h(r_N, \nu_{N, N}, r_N^j, \nu_{N, N}^j) + p(\mathbf{u}_{r, N}, -\mathbf{C}_{N, N}, \mathbf{u}_{r, N}^j) \leqslant 0,\\
			\sigma^2 - \nu_{N, N} \leqslant 0,
		\end{cases} \label{sub3-3} \\
		& p(\mathbf{u}_{r, k}, \mathbf{E}_{k, n}, \mathbf{u}_{r, k}^j) \leqslant 0, \begin{cases}
			k \in \{2,\cdots, N\},\\
			n \in \{1, \cdots, k-1\},
		\end{cases}\label{sub3-4}\\
		& \qquad\qquad\qquad\qquad\qquad (\ref{sub0-7}). \notag
	\end{align}
\end{subequations}

\subsection{Base-end MA Position Optimization}
Finally, we optimize the MA positions at the BS while keeping the beamforming parameters, power allocation, and user-end MA positions fixed. However, due to the strong coupling of the BS-end MA positions in the constraints, jointly optimizing all BS-end MA positions is highly complex. To address this, we apply the concept of AO algorithm: fixing all other MA positions, we optimize the position of one MA at a time. After determining the optimal position for this MA, we fix it and proceed to optimize the next MA position. This process continues iteratively until all BS-end MA positions are optimized. Consequently, we decompose the third subproblem into $M$ sub-subproblems.
\begin{subequations}
	\begin{align}
		\text{(P4.$m$.0)} &\qquad\quad\enspace \max_{\mathbf{u}_{t,m}, \mathbf{r}, \bm{\nu}} \left(\prod_{k=1}^{N}r_k\right)^{\frac{1}{N}}, \label{sub4.m.0-1}\\
		\text{s.t.} &\quad \Vert \mathbf{u}_{t,m} - \mathbf{u}_{t,l} \Vert_2 \geqslant \frac{1}{2}\lambda_c, l\in\mathcal{M}, l\neq m, \label{sub4.m.0-2}\\
		&\qquad\qquad\qquad \mathbf{u}_{t,m} \in \mathcal{C}_{t}, \label{sub4.m.0-3}\\
		&\qquad\qquad (\ref{sub1-2}), (\ref{sub1-3}), (\ref{sub1-4}).
	\end{align}
\end{subequations}

Similarly, the constraints in (P4.m.0) must be transformed to their equivalent or sufficiently convex forms. To begin, for (\ref{sub1-2}), we employ the same methodology as in the previous two subproblems: namely, constructing (\ref{equa-21}) as the global convex upper bound for the left-hand side of the first inequality constraint. For the right-hand side, we expand it into the following expression:
\begin{equation}\label{equa-43}
	\begin{split}
		\left|\mathbf{h}_k\mathbf{v}_n\right|^2 = \left|v_{n,m}\mathbf{f}_k\mathbf{\Sigma}_k\mathbf{g}_{k,m}+\sum_{j=1,j\neq m}^{M}v_{n,j}\mathbf{f}_k\mathbf{\Sigma}_k\mathbf{g}_{k,j}\right|^2.
	\end{split}
\end{equation}
To streamline the notation, we define $\bm{\eta}_{k}\triangleq\mathbf{f}_k\mathbf{\Sigma}_k$ and $t_{k,n}^m \triangleq \sum_{j=1,j\neq m}^{M}v_{n,j}\mathbf{f}_k\mathbf{\Sigma}_k\mathbf{g}_{k,j}$. Consequently, $\left|\mathbf{h}_{k}\mathbf{v}_{n}\right|$ can be further expressed as:
\begin{equation}\label{equa-44}
	\begin{split}
		&\left|\mathbf{h}_{k}\mathbf{v}_n\right|^2 = \vert v_{n,m}\vert^2\mathbf{g}_{k,m}^H\bm{\eta}_{k}^H\bm{\eta}_{k}\mathbf{g}_{k,m} + v_{n,m}^*t_{k,n}^m\mathbf{g}_{k,m}^H\bm{\eta}_{k}^H\\
		&\qquad\qquad\qquad + v_{n,m}(t_{k,n}^m)^*\bm{\eta}_{k}\mathbf{g}_{k,m} + \vert t_{k,n}^m\vert^2\\
		&\quad \triangleq \vert v_{n,m}\vert^2 \mathbf{g}_{k,m}^H \mathbf{F}_k \mathbf{g}_{k,m} + 2\Re\left(\mathbf{z}_{k,n}^m\mathbf{g}_{k,m}\right) + \vert t_{k,n}^m\vert^2,
	\end{split}
\end{equation}
where we define $\mathbf{F}_k \triangleq \bm{\eta}_{k}^H\bm{\eta}_{k}$ and $\mathbf{z}_{k,n}^m = v_{n,m}(t_{k,n}^m)^*\bm{\eta}_{k}$. Regarding the second inequality constraint of (\ref{sub1-2}), an analogous expansion of its right-hand side yields:
\begin{equation}\label{equa-45}
	\begin{split}
		&\sum_{j=n+1}^{N}\vert \mathbf{h}_k\mathbf{v}_j\vert^2 = \sum_{j=n+1}^{N}\vert v_{j,m}\vert^2\mathbf{g}_{k,m}^H\mathbf{F}_k\mathbf{g}_{k,m} \\
		&\qquad\qquad + 2\Re\left(\sum_{j=n+1}^{N}\mathbf{z}_{k,j}^m\mathbf{g}_{k,m}\right) + \sum_{j=n+1}^{N}\vert t_{k,j}^m\vert^2.
	\end{split}
\end{equation}

Given that (\ref{sub1-3}) constitutes a specific instance of (\ref{sub1-2}), the transformation applied to (\ref{sub1-3}) is virtually identical to that of (\ref{sub1-2}). For (\ref{sub1-4}), we begin by decomposing it into the form outlined in (\ref{equa-26}), subsequently rearranging and expanding the terms to derive the following:
\begin{equation}\label{equa-46}
	\begin{split}
		&\alpha\vert \mathbf{h}_k\mathbf{v}_{n+1}\vert^2 - \vert \mathbf{h}_k\mathbf{v}_{n}\vert^2 \\
		=& \left(\alpha\vert v_{n+1,m}\vert^2 - \vert v_{n, m}\vert^2  \right)\mathbf{g}_{k,m}^H \mathbf{F}_{k}\mathbf{g}_{k,m} + 2\Re\left[(-\mathbf{z}_{k,n}^m \right.\\
		&\left. +\alpha\mathbf{z}_{k,n+1}^m)\mathbf{g}_{k,m}\right] + \alpha\vert t_{k,n+1}^m \vert^2 - \vert t_{k, n}^m \vert^2.
	\end{split}
\end{equation}

An examination of (\ref{equa-44})-(\ref{equa-46}) reveals a recurrent structural pattern across these expressions. We isolate this common structure to formulate an auxiliary function, which serves to streamline the ensuing analysis. The constructed auxiliary function takes the form (\ref{equa-47}), where $\mathbf{O}\in\mathbb{C}^{L_t\times L_t}$ and $o_{i,j}$ is the $(i,j)$-th element of $\mathbf{O}$. $\bm{\tau} = [\tau_1, \cdots, \tau_{L_t}] \in \mathbb{C}^{1\times L_t}$ is an auxiliary complex vector and $l$ is a constant. The ability to generate (\ref{equa-44})-(\ref{equa-46}) through parameter variations in (\ref{equa-47}) implies that establishing a global convex upper bound for (\ref{equa-47}) will convert constraints (\ref{sub1-2})-(\ref{sub1-4}) into either equivalent or sufficient convex forms. 
\begin{figure*}[!t]
	\begin{align}
		\begin{split}
			T_k(\mathbf{u}_{t,m}, \mathbf{O}, \bm{\tau}, l) \triangleq \mathbf{g}_{k,m}^H\mathbf{O}\mathbf{g}_{k,m}+2\Re\left(\bm{\tau}\mathbf{g}_{k,m}\right) + l=& \sum_{i=1}^{L_t}\vert o_{i,i}\vert + 2\sum_{i=1}^{L_t-1}\sum_{j=i+1}^{L_t}\vert o_{i,j}\vert\cos\left[\frac{2\pi}{\lambda_c}(\rho_{k,m,j}^t-\rho_{k,m,i}^t) + \angle o_{i,j}\right]\\
			& + 2\sum_{i=1}^{L_t}\left(\Re(\tau_i)\cos\left(\frac{2\pi}{\lambda_c}\rho_{k,m,i}^t\right) - \Im(\tau_i)\sin\left(\frac{2\pi}{\lambda_c}\rho_{k,m,i}^t\right)\right) + l.
		\end{split}\label{equa-47}
	\end{align}
	\hrule
\end{figure*}

According to Taylor's theorem, for any given point $\mathbf{u}_{t,m}^j$, a global convex upper bound for $T_k(\mathbf{u}_{t,m}, \mathbf{O}, \bm{\tau}, l)$ can be constructed as:
\begin{equation}\label{equa-48}
	\begin{split}
		&t_k(\mathbf{u}_{t,m}, \mathbf{u}_{t,m}^j, \mathbf{O}, \bm{\tau}, l) \triangleq T_k(\mathbf{u}_{t,m}^j, \mathbf{O}, \bm{\tau}, l) \\
		&+ \triangledown T_k(\mathbf{u}_{t,m}^j, \mathbf{O}, \bm{\tau}, l)^T\left(\mathbf{u}_{t,m}-\mathbf{u}_{t,m}^j\right)^T + \frac{1}{2}\xi_{k,m}^{(\mathbf{O},\bm{\tau})}\\
		&\left(\mathbf{u}_{t,m}-\mathbf{u}_{t,m}^j\right)\left(\mathbf{u}_{t,m}-\mathbf{u}_{t,m}^j\right)^T.
	\end{split}
\end{equation}

To ensure feasibility, the value of $\xi_{k,m}^{(\mathbf{O},\bm{\tau})}$ must satisfy the condition $\xi_{k,m}^{(\mathbf{O},\bm{\tau})}\mathbf{I_{L_t}} \succeq \triangledown^2 T_k(\mathbf{u}_{t,m}, \mathbf{O}, \bm{\tau}, l)$. To determine the permissible values of $\xi_{k,m}^{(\mathbf{O},\bm{\tau})}$, we first compute the second-order partial derivatives of the function $T_k(\mathbf{u}_{t,m}, \mathbf{O}, \bm{\tau}, l)$ and identify that they have upper bounds, which are given by (\ref{equa-49}), (\ref{equa-50}), and (\ref{equa-51}). Given $\Vert \triangledown^2 T_k(\mathbf{u}_{t,m}, \mathbf{O}, \bm{\tau}, l)\Vert_F \geqslant \Vert \triangledown^2 T_k(\mathbf{u}_{t,m}, \mathbf{O}, \bm{\tau}, l)\Vert_2$, we can then determine that the admissible $\xi_{k,m}^{(\mathbf{O},\bm{\tau})}$ can be obtained in the following form:
\begin{figure*}[!t]
	\begin{equation}\label{equa-49}
		\resizebox{0.92\hsize}{!}{$
		\frac{\partial^2 T_k}{(\partial x_m^t)^2} \leqslant \frac{8\pi^2}{\lambda_c^2}\left[\sum_{i=1}^{L_t-1}\sum_{j=i+1}^{L_t}\vert o_{i,j}\vert(\cos\theta_{k,j}^t\cos\phi_{k,j}^t - \cos\theta_{k,i}^t\cos\phi_{k,i}^t )^2 + \sum_{i=1}^{L_t}(\vert \Re(\tau_i)\vert + \vert \Im(\tau_i)\vert)(\cos\theta_{k,i}^t\cos\phi_{k,i}^t)^2 \right] \triangleq\varsigma_{1,k,m}^{(\mathbf{O},\bm{\tau})}.
		$}
	\end{equation}
	\begin{equation}\label{equa-50}
		\resizebox{0.92\hsize}{!}{$
		\frac{\partial^2 T_k}{(\partial y_m^t)^2} \leqslant \frac{8\pi^2}{\lambda_c^2}\left[\sum_{i=1}^{L_t-1}\sum_{j=i+1}^{L_t}\vert o_{i,j}\vert(\cos\theta_{k,j}^t\sin\phi_{k,j}^t - \cos\theta_{k,i}^t\sin\phi_{k,i}^t )^2 + \sum_{i=1}^{L_t}(\vert \Re(\tau_i)\vert + \vert \Im(\tau_i)\vert)(\cos\theta_{k,i}^t\sin\phi_{k,i}^t)^2 \right] \triangleq\varsigma_{2,k,m}^{(\mathbf{O},\bm{\tau})}.
		$}
	\end{equation}
	\begin{equation}\label{equa-51}
		\begin{split}
			\resizebox{0.92\hsize}{!}{$
			\frac{\partial^2 T_k}{\partial x_m^t \partial y_m^t} \leqslant \frac{8\pi^2}{\lambda^2}\left[\sum_{i=1}^{L_t-1}\sum_{j=i+1}^{L_t}\vert o_{i,j}\vert\vert\cos\theta_{k,j}^t\cos\phi_{k,j}^t - \cos\theta_{k,i}^t\cos\phi_{k,i}^t\vert \vert \cos\theta_{k,j}^t\sin\phi_{k,j}^t - \cos\theta_{k,i}^t\sin\phi_{k,i}^t \vert + \sum_{i=1}^{L_t}(\vert \Re(\tau_i)\vert + \vert \Im(\tau_i)) \right.$} \\
			\resizebox{0.35\hsize}{!}{$ \left.  \left|\cos\theta_{k,i}^t\cos\phi_{k,i}^t\right| \left|\cos\theta_{k,i}^t\sin\phi_{k,i}^t\right|\right] \triangleq \varsigma_{3,k,m}^{(\mathbf{O},\bm{\tau})}.
			$}
		\end{split}
	\end{equation}
	\hrule
\end{figure*}
\begin{equation}\label{equa-52}
	\xi_{k,m}^{(\mathbf{O},\bm{\tau})} = \left( {\varsigma_{1,k,m}^{(\mathbf{O},\bm{\tau})}}^2 + {\varsigma_{2,k,m}^{(\mathbf{O},\bm{\tau})}}^2 + 2 {\varsigma_{3,k,m}^{(\mathbf{O},\bm{\tau})}}^2 \right)^{\frac{1}{2}}.
\end{equation}

Additionally, (\ref{sub4.m.0-2}) retains non-convexity in the variable $\mathbf{u}_{t,m}$, necessitating reformulation into either an equivalent or sufficiently convex constraint. Through the Cauchy-Schwarz inequality $\mathbf{a}^T \mathbf{b} \leqslant \Vert \mathbf{a}\Vert_2 \Vert \mathbf{b}\Vert_2, \mathbf{a}\in\mathbb{R}^{n}, \mathbf{b}\in \mathbb{R}^{n}$, given an arbitrary point $\mathbf{u}_{t,m}$, we derive the following inequality:
\begin{equation}\label{equa-53}
	\begin{split}
		\Vert \mathbf{u}_{t,m}^j - \mathbf{u}_{t,l}\Vert_2 \Vert \mathbf{u}_{t,m} - \mathbf{u}_{t,l}\Vert_2 \geqslant \left(\mathbf{u}_{t,m}^j - \mathbf{u}_{t,l}\right) \\
		\left(\mathbf{u}_{t,m} - \mathbf{u}_{t,l}\right), l\in \mathcal{M}, l\neq m \\
		\Longleftrightarrow \Vert \mathbf{u}_{t,m} - \mathbf{u}_{t,l}\Vert_2 \geqslant 
		\frac{\left(\mathbf{u}_{t,m}^j\ - \mathbf{u}_{t,l}\right) \left(\mathbf{u}_{t,m}-\mathbf{u}_{t,l}\right)}{\Vert \mathbf{u}_{t,m}^j - \mathbf{u}_{t,l}\Vert_2}\\
		,l\in\mathcal{M}, l\neq m.
	\end{split}
\end{equation}

Consequently, we establish the global concave lower bound for the left-hand side of (\ref{sub4.m.0-1}), whose substitution back into (\ref{sub4.m.0-1}) produces a sufficiently convex constraint formulation. Prior to converting (P4.m.0) into a tractable convex problem, we introduce the following auxiliary constants:

\begin{equation}\label{equa-54}
	\begin{split}
		&\qquad\quad \mathbf{I}_{k,n}^{(m)} \triangleq -\vert v_{n,m}\vert^2 \mathbf{F}_k, \quad \mathbf{J}_{k, n}^{(m)} \triangleq \sum_{j=n+1}^{N}\vert v_{j, m} \vert^2 \mathbf{F}_k,\\
		&\qquad\qquad\quad \mathbf{L}_{k,n}^{(m)} \triangleq \left(\alpha \vert v_{n+1, m}\vert^2 - \vert v_{n, m} \vert^2\right) \mathbf{F}_k;\\
		& \mathbf{d}_{k,n}^m \triangleq \sum_{j=n+1}^{N}\mathbf{z}_{k,j}^m,\enspace  \mathbf{e}_{k,n}^m \triangleq \alpha\mathbf{z}_{k, n+1}^m - \mathbf{z}_{k, n}^m;\enspace  i_{k,n}^m \triangleq -\vert t_{k,n}^m\vert^2, \\
		&\qquad\quad j_{k,n}^m \triangleq \sum_{j=n+1}^{N}\vert t_{k,j}^m\vert^2,\quad l_{k,n}^m \triangleq \alpha \vert t_{k,n+1}^m \vert^2 - \vert t_{k, n}^m \vert^2.
	\end{split}
\end{equation}

Having established these constants, we ultimately reformulate (P4.$m$.0) as the convex approximation (P4.$m$) with the following properties: (1) The feasible set of (P4.$m$) is contained within that of (P4.$m$.0); (2) Solutions optimal for (P4.$m$) preserve feasibility for (P4.$m$.0); (3) Solving (P4.$m$) provides at least sub-optimal solution for (P4.$m$.0). (P4.$m$) is expressed as:
\begin{subequations}
	\begin{align}
		&\text{(P4.$m$)} \qquad\qquad \max_{\mathbf{u}_{t,m}, \mathbf{r}, \bm{\nu}}\left(\prod_{k=1}^{N}r_k\right)^{\frac{1}{N}}, \label{sub4.m-1}\\
		&\text{s.t.} \begin{cases}
			\resizebox{0.9\hsize}{!}{$h(r_n, \nu_{k,n}, r_n^j, \nu_{k,n}^j) \leqslant - t_k(\mathbf{u}_{t,m}, \mathbf{u}_{t,m}^j, \mathbf{I}_{k,n}^{(m)}, -\mathbf{z}_{k,n}^m, i_{k,n}^m),$} \\
			t_k(\mathbf{u}_{t,m},\mathbf{u}_{t,m}^j,\mathbf{J}_{k,n}^{(m)}, \mathbf{d}_{k,n}^{(m)}, j_{k,n}^m) + \sigma^2 \leqslant \nu_{k,n},\\
			n \in \{1,\cdots, N-1\}, k\in\{n,\cdots,N\},
		\end{cases}\label{sub4.m-2}\\
		&\resizebox{0.92\hsize}{!}{$h(r_N, \nu_{N,N}, r_N^j, \nu_{N,N}^j) \leqslant - t_N(\mathbf{u}_{t,m}, \mathbf{u}_{t,m}^j, \mathbf{I}_{N,N}^{(m)}, -\mathbf{z}_{N,N}^m, i_{N,N}^m),$}\label{sub4.m-3}\\
		&\qquad\qquad\qquad\qquad  \sigma^2 - \nu_{N,N} \leqslant 0,\label{sub4.m-4}\\
		&\resizebox{0.85\hsize}{!}{$t_k(\mathbf{u}_{t,m},\mathbf{u}_{t,m}^{t},\mathbf{L}_{k,n}^{(m)},\mathbf{e}_{k,n}^m,l_{k,n}^m) \leqslant 0, \begin{cases}
			2\leqslant k \leqslant N,\\
			1\leqslant n \leqslant k-1,
		\end{cases} \label{sub4.m-5}$}\\
		& \frac{\left(\mathbf{u}_{t,m}^j-\mathbf{u}_{t,j}\right) \left(\mathbf{u}_{t,m}-\mathbf{u}_{t,j}\right)  }{\Vert_2 \mathbf{u}_{t,m}^j - \mathbf{u}_{t,j}\Vert_2} \geqslant \frac{1}{2}\lambda_c, \begin{cases}
			j\in \mathcal{M},\\
			j \neq m,
		\end{cases} \label{sub4.m-6}\\
		&\qquad\qquad\qquad\qquad\qquad (\ref{sub4.m.0-2}). \notag
	\end{align}
\end{subequations}

\subsection{Convergence and Complexity Analysis}

The procedural framework of the proposed optimization algorithm is detailed in \textbf{Algorithm \ref{algo-1}}. To jointly optimize the positions of MAs on both the user and BS sides, along with beamforming vectors and power allocation parameters, we design a nested-loop architecture, which comprises an outer loop that coordinates three dedicated inner loops, each addressing one of the following tasks: beamforming and power allocation, user-side MA positioning, and BS-side MA positioning. Within each inner loop, we employ the principles of SCA and SPCA to reformulate non-convex constraints into tractable convex forms.

Considering P2 as a representative case, the feasible set of (P2) is a subset of that of (P2.0), thereby ensuring that any solution optimal for (P2) also satisfies the feasibility conditions of (P2.0) and (P1). Let $\{ \hat{\mathbf{r}}, \hat{\bm{\lambda}}, \hat{\mathbf{V}} \}$ denote the optimal solution at the $j$-th iteration. By fixing this solution as the reference point $\{\mathbf{r}^{(j+1)}, \bm{\lambda}^{(j+1)}, \mathbf{V}^{(j+1)}\}$ for the $(j+1)$-th iteration, the point $\{ \hat{\mathbf{r}}, \hat{\bm{\lambda}}, \hat{\mathbf{V}} \}$ is still feasible for all constraints. Consequently, the objective value at iteration $(j+1)$ is guaranteed to be no less than that at iteration $j$, leading to a non-decreasing sequence of objective values throughout the optimization of (P2). This monotonic convergence property similarly extends to problems (P3) and (P4.m), owing to their analogous structures. Furthermore, the consistent inheritance of the initial point $\mathbf{r}$ across all subproblems ensures a globally non-decreasing objective sequence across the outer loop. Given that the feasible set of (P1) is closed and the objective function is continuous over this set, the overall algorithm is guaranteed to converge to a stationary, locally optimal solution.

On the other hand, the computational complexity of the algorithm is characterized by analyzing each subproblem individually. Denote the computational complexities of solving (P2), (P3), and (P4.m) as $\mathcal{O}( (2MN)^{3.5}\ln\frac{1}{\epsilon} ) \triangleq \mathcal{O}(O_1)$, $\mathcal{O}( \max(\frac{N^2}{2} + \frac{7N}{2}, \frac{3N^2}{2}+\frac{N}{2})^{3} (\frac{N^2}{2} + \frac{7N}{2})^{0.5} \ln\frac{1}{\epsilon} ) \triangleq \mathcal{O}(O_2)$, and $\mathcal{O}((M+N^2) N \ln\frac{1}{\epsilon})\triangleq \mathcal{O}(O_3)$, respectively. Suppose the outer loop requires $\chi$ iterations, and the expected iterations for the three inner loops are $\chi_1$, $\chi_2$, and $\chi_3$, respectively. Then the overall complexity of \textbf{Algorithm \ref{algo-1}} is given by: 
\begin{equation}\notag
	\mathcal{O}( \chi (\chi_1 O_1 + \chi_2 O_2 + \chi_3 M O_3) ).
\end{equation}

\begin{algorithm}
	\caption{Joint Beamforming, Power Allocation, and MAs Positioning Optimization}
	\label{algo-1}
	Find an inner point in the feasible set of (P1) as the initial point $\tilde{\mathbf{u}}_t^{(0)}, \tilde{\mathbf{u}}_r^{(0)}, \mathbf{V}^{(0)}, \mathbf{r}^{(0)}, \bm{\nu}^{(0)}$.
	
	Set maximal iterations $\vartheta, \vartheta_1, \vartheta_2, \vartheta_3$ for outer loop, solving (P2), (P3), and (P4.m) respectively.
	
	Initialize $i = 0$.
	
	\Repeat{convergence or $i >\vartheta$.}{
	Initialize $j = 0$, $\tilde{\mathbf{u}}_t = \tilde{\mathbf{u}}_t^{(i)}$, $\tilde{\mathbf{u}}_r = \tilde{\mathbf{u}}_r^{(i)}$, $\mathbf{V}^j = \mathbf{V}^{(i)}$, $\mathbf{r}^j=\mathbf{r}^{(i)}, \bm{\nu}^j=\bm{\nu}^{(i)}$.
	
	\Repeat{convergence or $j >\vartheta_1$.}{
	Set $\hat{\mathbf{r}}$, $\hat{\bm{\nu}}$, and $\tilde{\mathbf{V}}$ to be the optimal solution of (P2).
	
	Set $j\longleftarrow j+1.$
	
	Update $\mathbf{r}^j \longleftarrow \hat{\mathbf{r}}$, $\bm{\nu}^j \longleftarrow \hat{\bm{\nu}}$, and $\mathbf{V}^j \longleftarrow \tilde{\mathbf{V}}$.
	}
	
	Update $\mathbf{V}^{(i+1)} \longleftarrow \mathbf{V}^j$, $\mathbf{r} \longleftarrow \mathbf{r}^j$, $\bm{\nu} \longleftarrow \bm{\nu}^j$.
	
	Initialize $j=0$, $\tilde{\mathbf{u}}_r^j = \tilde{\mathbf{u}}_r^{(i)}$, $\mathbf{V}=\mathbf{V}^{(i+1)}$,$\mathbf{r}^j=\mathbf{r}$, $\bm{\nu}^j=\bm{\nu}$.
	
	\Repeat{convergence or $j > \vartheta_2$}{
	Set $\hat{\mathbf{r}}$, $\hat{\bm{\nu}}$, and $\hat{\mathbf{u}}_r$ to be the optimal solution of (P3).
	
	Set $j \longleftarrow j+1$.
	
	Update $\mathbf{r}^j \longleftarrow \hat{\mathbf{r}}$, $\bm{\nu}^j \longleftarrow \hat{\bm{\nu}}$, and $\tilde{\mathbf{u}}_r^j \longleftarrow \hat{\mathbf{u}}_r$.
	}
	Update $\tilde{\mathbf{u}}_r^{(i+1)} \longleftarrow \tilde{\mathbf{u}}_r^j$, $\mathbf{r} \longleftarrow \mathbf{r}^j$, $\bm{\nu} \longleftarrow \bm{\nu}^j$.
	
	Initialize $\tilde{\mathbf{u}}_r \longleftarrow \tilde{\mathbf{u}}_r^{(i+1)}$.
	
	\For{m = 1 to M}{
	Initialize $j = 0, \mathbf{u}_{t,m}^j = \mathbf{u}_{t,m}$, $\mathbf{r}^j=\mathbf{r}$, $\bm{\nu}^j=\bm{\nu}$.
	
	\Repeat{convergence or $j > \vartheta_3$}{
	Set $\hat{\mathbf{r}}$, $\hat{\bm{\nu}}$, and $\hat{\mathbf{u}}_{t,m}$ to be the optimal solution of (P4.m).
	
	Set $j \longleftarrow j+1$.
	
	Updata $\mathbf{r}^j \longleftarrow \hat{\mathbf{r}}$, $\bm{\nu}^j \longleftarrow \hat{\bm{\nu}}$, and $\mathbf{u}_{t,m}^j \longleftarrow \hat{\mathbf{u}}_{t,m}$.
	}
	
	Updata $\mathbf{u}_{t,m} \longleftarrow \mathbf{u}_{t,m}^j$, $\mathbf{r} \longleftarrow \mathbf{r}^j$, $\bm{\nu} \longleftarrow \bm{\nu}^j$.
	}
	
	Updata $\tilde{\mathbf{u}}_t^{(i+1)} \longleftarrow \tilde{\mathbf{u}}_t$, $\mathbf{r}^{(i+1)} \longleftarrow \mathbf{r}$, $\bm{\nu}^{(i+1)} \longleftarrow \bm{\nu}$.
	
	Set $i \longleftarrow i+1$.
	}
	
	\Return{$\tilde{\mathbf{u}}_t^{(i)}$, $\tilde{\mathbf{u}}_r^{(i)}$, $\mathbf{V}^{(i)}$}.
\end{algorithm}

\section{Simulation Results} \label{section4}
In this section, we describe the simulation setup and present the corresponding numerical results, which validate the effectiveness of the proposed MA-assisted NOMA downlink multiuser wireless communication system and its optimization algorithm.

\subsection{Simulation Setup}
In the default simulation setup, we consider a downlink wireless communication system in which a BS equipped with $M=16$ MAs serves $N=4$ users. Each user is randomly located within a ring-shaped area centered at the BS, with a distance uniformly distributed as $d_k \sim \mathcal{U}[50, 200]$ meters. The carrier wavelength is set to $\lambda_c = 0.01$m, and the maximum transmit power of the BS is $P = 30$dBm. The BS-user channels follow a geometric propagation model, implying that the transmit and receive paths are reciprocal and one-to-one, i.e., $L_k^t=L_k^r=L=10, \forall k \in \mathcal{N}$. For the $k$-th user, the diagonal elements of its PRM are assumed to follow a CSCG distribution, i.e., $\mathcal{CN}(O, c_0d_k^{-\alpha_0}L^{-1})$, where $c_0 = \frac{G_tG_r\lambda_c^2}{(4\pi)^2}$ represents the average channel gain at a reference distance of 1 meter; the transmit and receive antenna gains are both set to $G_t=G_r=1$, and $\alpha_0=2.8$ denotes the path loss factor. The power of the AWGN at the receiver is $\sigma^2 = -80$dBm. The transmit angle and receive angle for each user $k$ are both assumed to follow uniform distributions, i.e., $\theta_{k,i}^t, \theta_{k, i}^r \sim \mathcal{U}[0, \frac{\pi}{2}], \forall k \in \mathcal{N}, \forall i \in \{1, \cdots , L\}$, and $\phi_{k, j}^t, \phi_{k,j}^r \sim \mathcal{U}[0, 2\pi], \forall k \in \mathcal{N}, \forall j \in \{1, \cdots , L\}$. The movable regions for the MAs at the BS and at the users are $R_t = 20\lambda_c$ and $R_r = 4\lambda_c$.

\subsection{Convergence Performance of Proposed Algorithm}

\begin{figure}[!h]
	\centering
	\includegraphics[width=1\linewidth]{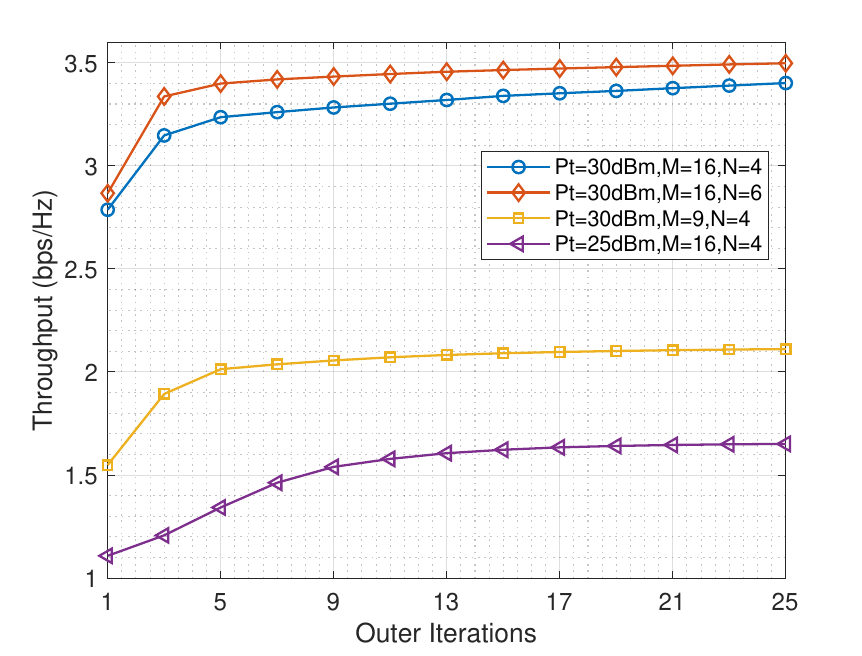}
	\caption{Convergence Performance.}
	\label{convergence}
\end{figure}

Fig. \ref{convergence} illustrates the convergence behavior of the proposed optimization algorithm for the MA-assisted downlink NOMA multiuser wireless communication system. To evaluate its effectiveness, the algorithm was applied to multiple scenarios, each yielding distinct optimization trajectories. The results consistently demonstrate that, irrespective of variations in the base station's transmit power threshold, the number of antennas, or the number of users, the algorithm produces a monotonically increasing sequence of system throughput that converges reliably. Moreover, in all considered scenarios, convergence is achieved within 0.01 bps/Hz after approximately 15 outer-loop iterations.

\subsection{Performance Comparison with Benchmark Schemes}
We refer to the proposed MA-assisted NOMA downlink multiuser scenario as NOMA-MA. To clearly illustrate the throughput gains achieved by this scheme, we compare it with six benchmark approaches, namely: NOMA-MA(UE), NOMA-FPA, SDMA-MA, SDMA-FPA, TDMA-MA, and TDMA-FPA. The detailed descriptions of each scheme are provided below.

\begin{itemize}
	\item[1)] \textbf{NOMA-MA:} Our proposed MA-enhanced NOMA downlink multiuser system deploys MAs at both BS and user sides. Through simultaneous optimization of BS beamforming vectors, Power allocation parameters, and, MA positioning at both ends, the system achieves maximal throughput performance.
	
	\item[2)] \textbf{NOMA-FPA:} This configuration also adopts NOMA, employing a uniformly spaced fixed antenna array on the BS plane and fixed antennas located at the origin of each user’s plane.
	
	\item[3)] \textbf{SDMA-MA:} Adopting SDMA for multiple access, each user receives the composite signal transmitted by the BS, treating the signals intended for other users as interference. Both the BS and users are equipped with MAs. A modified zero-forcing (ZF)-based algorithm is employed for beamforming optimization.
	
	\item[4)] \textbf{SDMA-FPA:} Similarly, in the SDMA-based multiple access configuration, both the BS and users are equipped with FPAs. The BS antennas are uniformly distributed across the BS plane, while the user antennas are fixed at the origin of each user's plane.
	
	\item[5)] \textbf{TDMA-MA:} This scheme employs MA configurations at both the BS and user ends, utilizing TDMA, where the BS sequentially serves users within designated time slots.MA positions are updated during the transitions between users. System throughput is maximized through joint optimization of MA positioning, time slot allocation, and fairness-aware user experience balancing.
	
	\item[6)] \textbf{TDMA-FPA:} Also adopting TDMA, both the BS and users are equipped with FPAs, consistent with the previously described FPA schemes.
\end{itemize}

\begin{figure*}[!t]
	\centering
	\begin{minipage}{0.32\linewidth}
		\centering
		\includegraphics[width=1\linewidth]{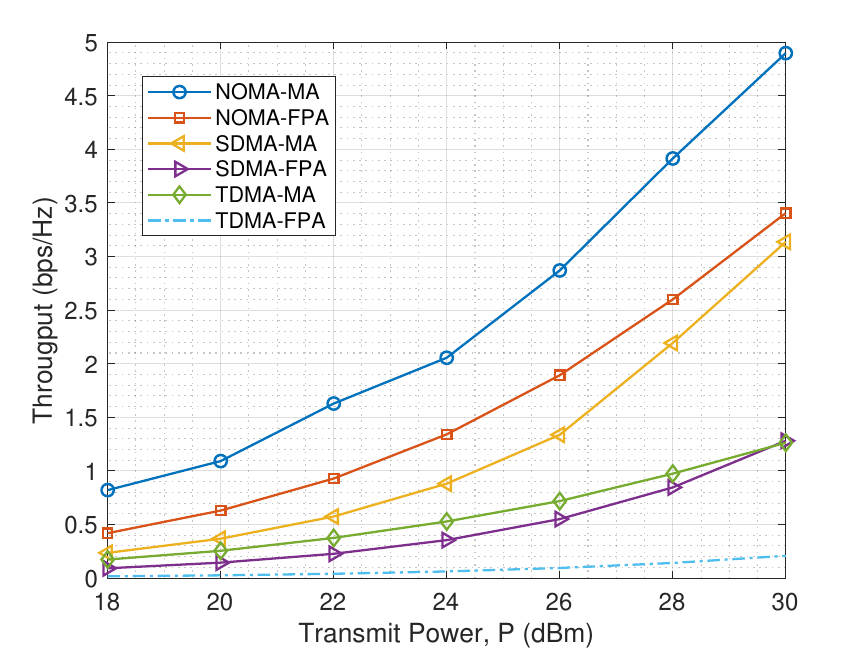}
		\caption{Throughput versus the transmit power threshold.}
		\label{power}
	\end{minipage}
	\begin{minipage}{0.32\linewidth}
		\centering
		\includegraphics[width=1\linewidth]{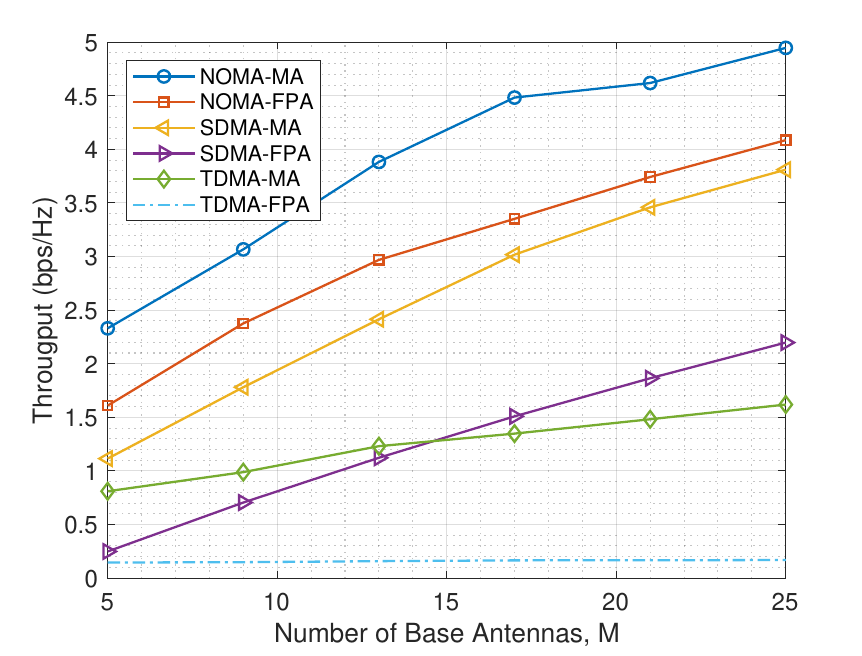}
		\caption{Throughput versus the number of antennas at BS.}
		\label{antenna}
	\end{minipage}
	\begin{minipage}{0.32\linewidth}
		\centering
		\includegraphics[width=1\linewidth]{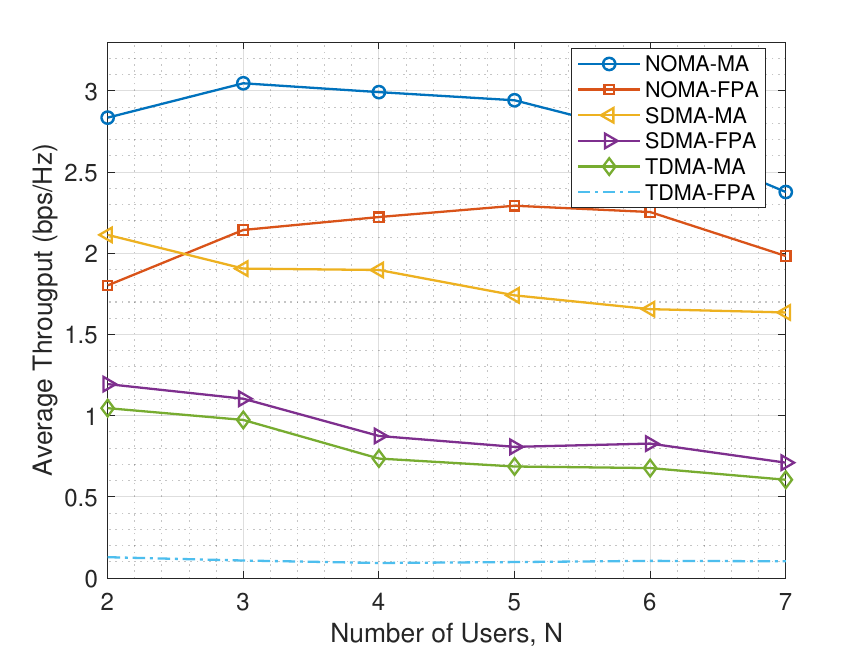}
		\caption{Average throughput versus the number of users.}
		\label{users}
	\end{minipage}
\end{figure*}

\begin{figure*}
	\centering
	\begin{minipage}{0.32\linewidth}
		\centering
		\includegraphics[width=1\linewidth]{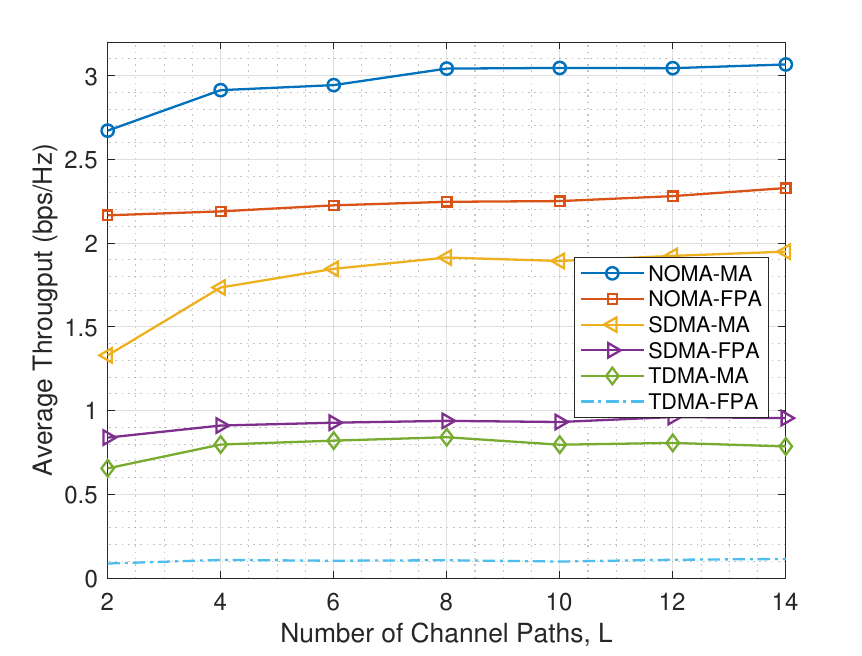}
		\caption{Average throughput versus the number of channel paths.}
		\label{paths}
	\end{minipage}
	\begin{minipage}{0.32\linewidth}
		\centering
		\includegraphics[width=1\linewidth]{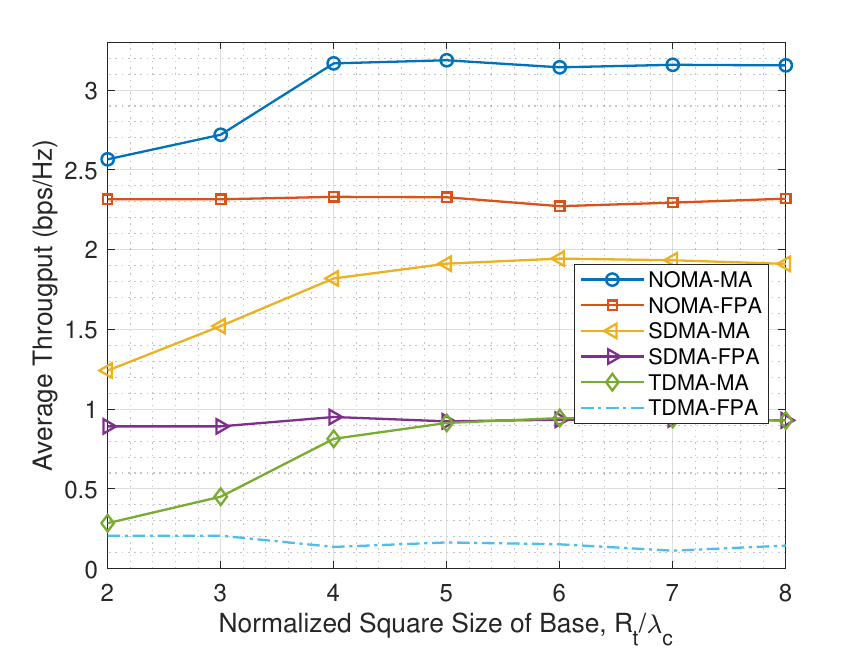}
		\caption{Average throughput versus the movable region size at BS.}
		\label{size_base}
	\end{minipage}
	\begin{minipage}{0.32\linewidth}
		\centering
		\includegraphics[width=1\linewidth]{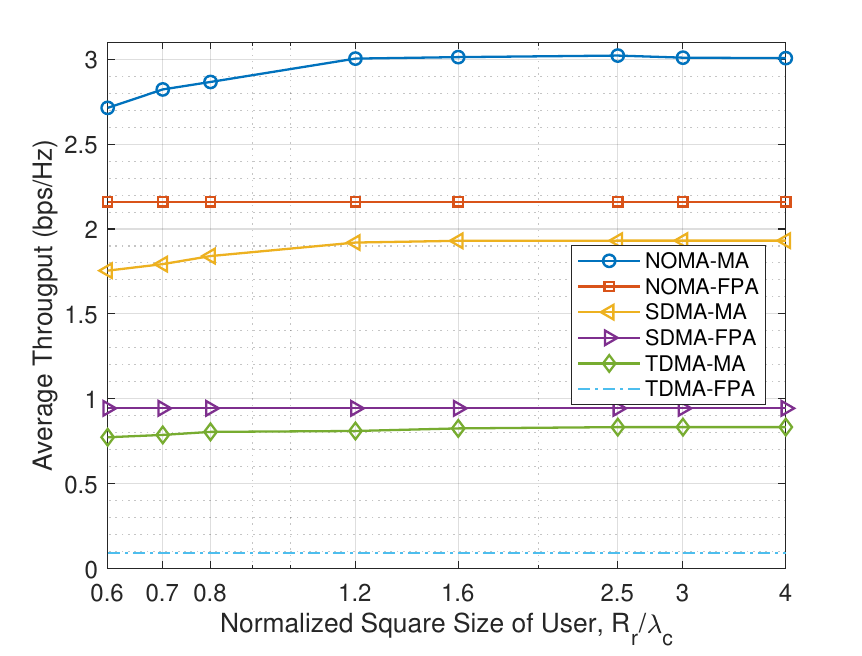}
		\caption{Average throughput versus the movable region size at user side.}
		\label{size_user}
	\end{minipage}
\end{figure*}

Fig. \ref{power} illustrates the system throughput versus the transmit power threshold at the BS for different scenarios. As expected, all configurations exhibit a monotonically increasing throughput with rising power budgets, indicating a clear positive correlation. Among them, the proposed NOMA-MA consistently outperforms all other schemes across the entire power range. An interesting observation arises from the comparison between the SDMA-FPA and TDMA-MA. At lower power thresholds, TDMA-MA achieves higher throughput, primarily due to its ability to allocate full transmit power to each user during dedicated time slots, along with the channel enhancement enabled by MA positioning. However, as the threshold increases, SDMA-FPA gradually surpasses TDMA-MA. This crossover occurs because, at higher power levels, the SINR  is no longer a limiting factor for SDMA-FPA, allowing its superior spatial resource utilization to dominate. In contrast, the performance gains from MA optimization in TDMA-MA diminish in high-power regimes, whereas SDMA's spatial multiplexing capabilities become increasingly advantageous.

Fig. \ref{antenna} illustrates the relationship between throughput and the number of antenna elements at the BS. The results indicate that throughput increases across all schemes as the number of BS antennas grows, primarily due to the enhanced beamforming precision afforded by additional antenna elements, which improves signal focusing toward users. Notably, the comparative analysis shows that SDMA-FPA eventually outperforms TDMA-MA as the antenna count increases. This crossover is driven by SDMA’s improved ability to spatially separate user signals with expanded antenna arrays. Overall, the proposed NOMA-MA scheme consistently achieves the highest system throughput across all evaluated antenna configurations, underscoring its superior performance.

Fig. \ref{users} presents the average system throughput versus the number of served users. All NOMA-based schemes exhibit a characteristic trend: average throughput initially increases with the number of users but subsequently declines. This pattern reflects a fundamental trade-off in NOMA systems. While adding users introduces additional data streams that can enhance overall throughput, it also increases interference, thereby degrading the achievable rates for existing users. The throughput peaks at the point where the marginal gain from additional users exceeds the aggregate rate loss of current users; beyond this point, the performance deteriorates. Notably, the proposed NOMA-MA architecture consistently achieves superior throughput across all user population sizes.

Fig. \ref{paths} demonstrates the relationship between average system throughput and the number of channel paths under multiple channel realizations. The results indicate that an increase in channel paths enhances channel diversity, thereby improving throughput across all schemes. Notably, all MA-assisted scenarios exhibit significantly higher throughput gains with increasing channel paths compared to their FPA counterparts. This improvement is attributed to the MA systems’ ability to optimize antenna positioning, enabling more effective exploitation of multipath diversity. These findings confirm that MA systems outperform FPA configurations in throughput performance by leveraging multipath propagation more efficiently.

Fig. \ref{size_base} depicts the relationship between average system throughput and the movable region size of the BS MAs. The results indicate that expanding the BS MA movable region enhances system throughput by offering greater positioning flexibility, which facilitates more effective channel gain optimization and mitigates inter-antenna interference. In all MA scenarios, throughput increases monotonically with the movable region size until reaching a saturation point, beyond which further expansion yields negligible improvements. Conversely, FPA scenarios exhibit constant throughput regardless of the BS plane size, owing to their static and uniformly distributed antenna arrays.

Fig. \ref{size_user} illustrates the average system throughput versus the movable region size of user MAs. The results indicate that expanding the user MA movable region enhances throughput across all MA-enabled user configurations, whereas fixed-antenna user scenarios exhibit constant throughput, in line with theoretical expectations. However, the throughput gains exhibit diminishing returns beyond certain scenario-specific thresholds. Notably, NOMA-MA and SDMA-MA achieve greater throughput enhancement with expanded user-end MA regions. In contrast, TDMA-MA shows only marginal throughput improvements with larger user MA regions, as its interference-free nature reduces the potential benefits of user MA mobility; near-optimal channel gain can be achieved through BS MA positioning alone. Among all schemes, the proposed NOMA-MA consistently delivers the highest throughput across the entire range of movable region sizes.

\begin{figure}
	\centering
	\includegraphics[width=1\linewidth]{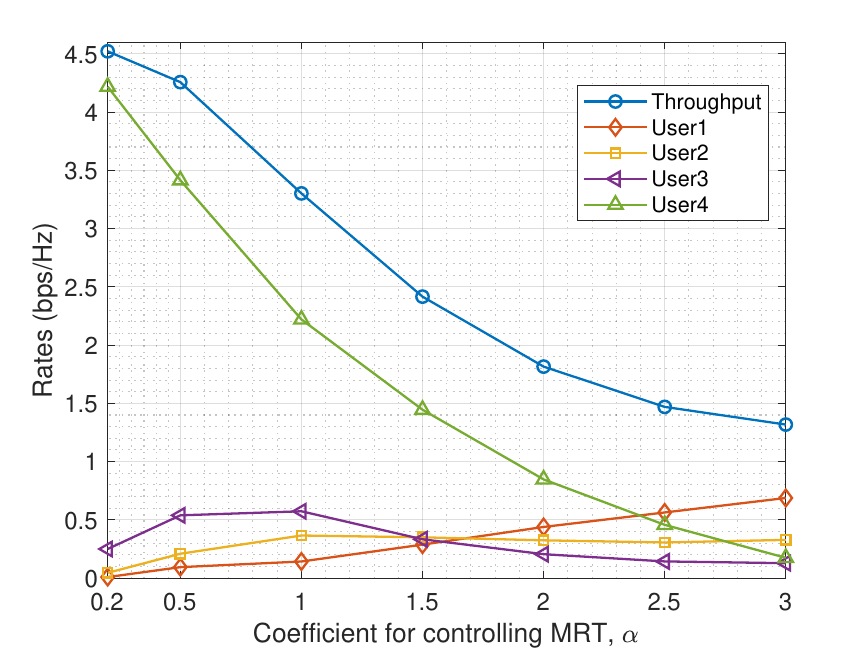}
	\caption{Impact of coefficient for MRT controlling.}
	\label{coe_MRT}
\end{figure}

For users with lower SIC decoding orders, their maximum achievable rates depend not only on their own decoding capabilities but also on the ability of subsequent users to successfully cancel interference. The introduction of the parameter $\alpha$ serves to regulate the rate at which other users can perform this cancellation. Simultaneously, for users with lower SIC decoding orders, the $\alpha$ constraint imposes an upper bound on the transmit power that can be allocated to them. Fig. \ref{coe_MRT} illustrates the relationship between system throughput and each user’s maximum achievable rate under different $\alpha$ values. As $\alpha$ increases, the achievable rate of the user closest to the BS decreases, indicating that power allocation to the user with the strongest channel is increasingly restricted by the optimization. Conversely, the achievable rate of the user farthest from the BS continues to rise. For intermediate users, their achievable rates initially increase and then decline. This trend arises because, at moderate $\alpha$  values, the stricter constraint on the closest user redistributes resources to intermediate users. However, as $\alpha$ increases further, these intermediate users begin to act as bottlenecks for users farther away, resulting in constrained allocations.

\section{Conclusion} \label{section5}
This paper investigated the joint optimization of beamforming, power allocation, and MA positioning at both the BS and user sides to maximize system throughput in downlink NOMA multiuser wireless communication systems with a predetermined SIC decoding order. We first established a channel model for this scenario and formulated a system throughput maximization problem. To address the non-convexity inherent in both the objective function and the constraints, an efficient approximation-based algorithm was developed by integrating techniques from the SCA and SPCA frameworks. This algorithm is capable of achieving a stable, locally optimal solution. Numerical simulations were conducted to evaluate the proposed system, and the results demonstrate that our MA-assisted downlink NOMA system significantly outperforms benchmark schemes in terms of throughput, thereby validating the effectiveness of the proposed optimization strategy. A promising direction for future research involves the optimization of the SIC decoding order within MA-enabled NOMA systems, which remains an open and compelling area for investigation.

\end{document}